\documentclass[eqsecnum,showpacs,showkeys,prd]{revtex4}
\usepackage{amsmath}
\begin{document}

\title{Quantum-Mechanical Model of Spacetime}

\author{Jarmo M\"akel\"a}
\email[Electronic address: ]{jarmo.makela@puv.fi} 
\affiliation{Vaasa University of Applied Sciences,
Wolffintie 30, 65200 Vaasa, Finland}

\begin{abstract}
We consider a possibility to construct a quantum mechanical model of spacetime, where 
Planck size quantum black holes act as the fundamental constituents of space and time. 
Spacetime is assumed to be a graph, where black holes lie on the vertices. Our model 
implies that area has a discrete spectrum with an equal spacing. At macroscopic length scales our model reproduces Einstein's
field equation with a vanishing cosmological constant as a sort of thermodynamical equation
of state of spacetime and matter fields. In the low temperature limit, where most black 
holes are assumed to be in the ground state, our model implies the Unruh and the Hawking
effects, whereas in the high temperature limit we find, among other things, that black hole
entropy depends logarithmically on the event horizon area, instead of being proportional
to the area. 
\end{abstract}



\maketitle

\section{Introduction}

      Gravitation is the most interesting of all known fundamental interactions of nature.
Its main interest lies in the fact that the properties of gravitation are closely linked 
to the properties of space and time. Indeed, we have learned from Einstein's general theory 
of relativity that gravitation is interaction between matter fields and the geometry of 
spacetime. Because of the intimate relationship between gravitation, space and time it is
unavoidable that any plausible theory of gravity always involves a theory of space and time.

    In this paper we consider a modest proposal for a model of spacetime which takes into 
account the quantum effects on its structure and properties. In other 
words, we consider a quantum mechanical model of spacetime. It is generally believed that 
quantum effects of gravity will dominate at length scales which are characterized by the so 
called Planck length
\begin{equation}
\ell_{Pl} := \sqrt{\frac{\hbar G}{c^3}} \approx 1.6\times 10^{-35}m.
\end{equation}
By means of our model we consider the structure and the properties of spacetime at the 
Planck length scale, and how classical spacetime as such as we know it from Einstein's 
general relativity emerges out from quantum spacetime at macroscopic length scales. An 
advantage of our model is that it reproduces, in a fairly simple manner, the "hard facts" 
of gravitational physics as such as we know them today. For instance, at macroscopic length 
scales our model reproduces Einstein's field equation with a vanishing cosmological 
constant. In the semiclassical limit our model reproduces, for low temperatures, the 
Hawking and the Unruh effects. Our model also makes some novel predictions, which
have already been anticipated, on various grounds, by other authors. Among other things
our model predicts that area has a discrete spectrum with an equal spacing, and that in
high enough temperatures black hole entropy depends logarithmically on the event
horizon area, instead of being proportional to the area.

    The basic idea of our model may be traced back to Jacobson's very important discovery
of the year 1995 that Einstein's field equation may actually be understood as a sort of
thermodynamical equation of state of spacetime and matter fields \cite{yksi, lisayksi}. More precisely, Jacobson 
considered the flow of heat through the local past Rindler horizon of an accelerated 
observer, and he {\it assumed} that when heat flows through a finite part of the horizon, 
then the amount of entopy carried through that part is, in natural units, one-quarter of 
the decrease in the area of that part. Identifying the Unruh temperature of the 
accelerated observer as the temperature assigned to the heat flow Jacobson found, by 
using the Raychaudhuri equation, that Einstein's field equation is a straightforward 
consequence of the fundamental thermodynamical relation
\begin{equation}
\delta Q = T\,dS
\end{equation}
between the heat $\delta Q$, absolute temperature $T$, and entropy $dS$ carried through 
the horizon. So it appears that Eq.(1.2), which was first introduced by Clausius in 1865
as a thermodynamical definition of the concept of entropy \cite{kaksi}, is not only one of 
the most important equations of thermodynamics, but it is also the fundamental equation of
gravitation. When horizon area is interpreted as a measure of entropy associated with 
the horizon, classical general relativity with all of its predictions is just one of the
consequences of Eq.(1.2). The close relationship between gravitation and thermodynamics was 
already suggested by the Bekenstein-Hawking entropy law which states that black hole has 
entropy $S_{BH}$ which, in natural units, is one-quarter of its horizon area or, in SI 
units \cite{kolme, nelja},
\begin{equation}
S_{BH} = \frac{1}{4}\frac{k_Bc^3}{\hbar G}A,
\end{equation}
but Jacobson's discovery makes the case clear: On the fundamental level classical gravity 
{\it is} thermodynamics of spacetime and matter fields \cite{viisi}.

    If the thermodynamical interpretation of gravity turns out to be correct, the problem of 
quantum gravity, or a theory which would bring together quantum mechanics and general relativity,
becomes clear: Instead of attempting to quantize Einstein's classical general relativity as 
if it were an ordinary field theory, we should try to identify the fundamental, microscopic
constituents of spacetime, and to postulate their quantum mechanical and statistical 
properties. These fundamental constituents play in the fabric of spacetime a role similar to that of 
atoms in the fabric of matter. Presumably those constituents are Planck size objects. 
The thermodynamical properties of spacetime follow from the statistics of its constituents,
and at appropriate length scales those thermodynamical properties should reproduce, at 
least in low temperatures, the well known effects of classical and semiclassical gravity. 
Although such thoughts were already expressed by Misner, Thorne and Wheeler in the inspired 
final chapter of their book \cite{kuusi}, it is curious that the approach described above has never, to 
the best knowledge of the author, been attempeted systematically. Loop quantum gravity bears
some resemblance to such an approach, but its roots lie in the attempts to quantize 
Einstein's general relativity as if it were an ordinary field theory, and its statistical 
and thermodynamical aspects are still unknown \cite{seitseman}.

      Our model may be regarded as an attempt to begin to fill in this gap in the existing 
literature of gravitational physics. To avoid any possible misunderstandings it should be 
stressed that our model should {\it not} be considered as a proposal for a quantum theory
of gravity, or anything of the kind. Rather, it is an attempt to show how the "hard facts" 
of classical and semiclassical gravity may be obtained from very simple postulates 
concerning the properties of spacetime at the Planck length scales. In other words, it is
hoped that the role of our model would be that of encouragement: In contrast to the common
beliefs the correct theory of quantum gravity may ultimately turn out fairly {\it simple}.
After all, general relativity has now been with us for 90 years, and quantum mechanics for
80 years. The conceptual and the mathematical details of these theories have been analyzed
thoroughly, and as scientific theories they bear no mysteries whatsoever. Actually, it has 
turned out that they are both fairly simple and easy to understand. A combination of two 
theories which are both simple and easy to understand is not necessarily complicated and 
difficult to understand. 

     The first problem is to identify the fundamental constituents of spacetime. At the 
moment we have no idea what these fundamental constituents might actually be, but some 
hints may be gained by means of some general, quantum mechanical arguments. For instance,
suppose that we have closed a particle inside a box whose edge length is one Planck length
$\ell_{Pl}$. In that case it follows from Heisenberg's uncertainty principle that the 
momentum of the particle has an uncertainty $\Delta p\sim \hbar/\ell_{Pl}$. In the 
ultrarelativistic limit this uncertainty in the momentum corresponds to the uncertainty 
$\Delta E\sim c\Delta p$ in the energy of the particle. In other words, inside a box with
an edge length equal to $\ell_{Pl}$ we have closed a particle whose uncertainty in its 
energy is about the same as the so called Planck energy
\begin{equation}
E_{Pl} := \sqrt{\frac{\hbar c^5}{G}}\approx 2.0\times 10^9 J.
\end{equation}
This amount of energy, however, is enough to shrink the spacetime region surrounding the 
cube into a {\it black hole} with a Schwarzschild radius equal to around one Planck length.
So it is possible that one encounters with Planck size black holes when probing spacetime at
the Planck length scale. Such an idea is far from new. For instance, Misner, Thorne and 
Wheeler write in their book, how "[gravitational] collapse at the Planck scale of distance
is taking place everywhere and all the time in the quantum fluctuations in the geometry and,
one believes, the topology of spacetime."\cite{kuusi} These kind of sentences immediately bring into 
one's mind mental images of wormholes and tiny black holes furiously bubbling as a sort of
spacetime foam. Unfortunately, the idea of spacetime being made of Planck size black holes 
has never been taken very far \cite{kahdeksan}.

        In our model we take most seriously the idea that spacetime might really be made 
of Planck size black holes. In other words, we consider Planck size quantum black holes 
as the fundamental constituents of space and time. We postulate certain properties for 
such black holes, and from these postulates we deduce the macroscopic properties of space
and time. 

     We begin the construction of our model in Section 2 by defining the concept of a 
microscopic quantum black hole. The fundamental, undefined quantity associated with a 
microscopic quantum black hole is its {\it horizon area}. Following a proposal already 
made by Bekenstein in 1974 \cite{yhdeksan} we shall assume that the spectrum of the horizon area operator
is discrete with an equal spacing. Horizon area is the only observable associated with a 
microscopic black hole, and all properties of space and time
may ultimately be reduced back to the horizon area eigenvalues of the Planck size black 
holes constituting spacetime. Other possible quantities associated with black holes,
such as the mass, for instance, have no relevance at the Planck scale of distance. Actually,
even the concept of distance is abandoned. The fundamental concept is horizon area, and the
concept of distance, as well as the other metric concepts associated with spacetime arise 
as sort of statistical concepts at macroscopic length scales. 

  In Section 3 we construct a precise definition of spacetime. Our definition is based on 
the concept of {\it graph}, which is one of the simplest concepts of mathematics. In very 
broad terms, we postulate that spacetime is a certain kind of graph, where black
holes lie on the vertices. Our definition of spacetime is constructed in such a way that 
it makes possible to define the concept of {\it spacelike two-graph}, which is an analogue 
of the concept of {\it spacelike two-surface} in classical spacetime. At this stage there 
is no notion of time in spacetime, but the word "spacelike" is just a name.
   
     In Section 4 we proceed to the statistics of spacetime. At this stage spacelike 
two-graphs take a preferred role. We express four {\it independence}- and two 
{\it statistical} 
postulates for the black holes lying on the vertices of a spacelike two-graph. Among other
things our postulates state, in very broad terms, that the area of a spacelike two-graph
is proportional to the sum of the horizon areas of the black holes lying on its vertices,
and that the macroscopic state of a spacelike two-graph is determined by its area, whereas
its microscopic states are determined by the combinations of the horizon area eigenstates
of the holes. In other words, area takes the role of energy as a fundamental statistical 
quantity in our model in a sense that the microscopic and the macroscopic states of a 
spacelike two-graph are labelled by the horizon areas, instead of the energies, of its 
constituents. Our postulates imply that in the so called low temperature limit, where 
most black holes are on their ground states, the natural logarithm of the number of 
microstates corresponding to the same macroscopic state of the spacelike two-graph is 
directly proportional to the area of that graph. This is a very important conclusion, and
the "hard facts"of the gravitational physics are, in the low temperature limit, its 
simple consequences.

      In Section 5 we consider how classical spacetime as such as we know it from Einstein's
general theory of relativity emerges from quantum spacetime at macroscopic length 
scales. When worked out with a mathematical precision, the process of reconstruction of 
classical out of quantum spacetime is a rather long one. However, the basic idea may be 
summarized, again in very broad terms, by saying that at macroscopic length scales for every
vertex $w$ of spacetime there exists a subgraph $G_w$ in such a way that $G_w$ may be 
approximated, in a certain very specific sense, by a geometrical four-simplex $\sigma_w$ 
such that the areas of the two-faces, or triangles, of the four-simplex $\sigma_w$ are 
determined by the horizon areas of the black holes in $G_w$. In our model the geometrical 
four-simplex $\sigma_w$ plays the role analogous to that of tangent space in classical
general relativity. Every four-simplex $\sigma_w$ is equipped with a flat Minkowski metric.
It is a specific property of four-simplices that each four-simplex has an equal number of 
triangles and edges. Because of that the lengths of edges may be expressed in terms of the 
areas of triangles which, moreover, are determined by the horizon areas of the Planck size 
black holes. In other words, the metric properties of spacetime may be reduced to the 
horizon area eigenstates of its quantum constituents. However, it should be noted that the 
metric of spacetime is a statistical quantity, which is meaningful at macroscopic length 
scales only. After introducing the concept of metric in spacetime, it is possible to define other
concepts familiar from classical general relativity, such as the Christoffel symbols, the 
Riemann and the Ricci tensors, and so forth.

     In Section 6 we proceed to the thermodynamics of spacetime. As a simple generalization
of the concept of energy as such as it is defined in stationary spacetimes by means of the 
so called Komar integrals, we define the concept of {\it heat change} of a spacelike 
two-surface of spacetime. We introduce a specific picture of the propagation of radiation
through spacetime, where radiation propagates through a succession of spacelike two-surfaces 
of spacetime such that when radiation propagates through a spacelike two-surface, it picks 
up heat and entropy from the two-surface. Using this picture and our definition of the 
concept of heat change we find 
that the Unruh effect is a simple and straigthforward consequence of the result of Section
4 that, in the low temperature limit, every spacelike two-graph possesses an entropy, which
is proportional to its area. More precisely, we find that every observer in a uniformly
accelerating motion will observe thermal radiation with a temperature, which is proportional
to the proper acceleration of the observer. At macroscopic length scales spacelike 
two-graphs become spacelike two-surfaces, and a comparison of our expression for 
temperature to the Unruh temperature will fix the constant of proportionality such that in
the low temperature limit the entropy associated with every spacelike two-surface of 
spacetime is, in natural units, exactly {\it one-half} of its area. This result is closely
related to the Bekenstein-Hawking entropy law of Eq.(1.3). We shall also see that the 
Hawking effect is a straigthforward consequence of our result concerning the relationship 
between area and entropy of a spacelike two-surface.

     In Section 7 we derive Einstein's field equation from our model. Our derivation bears
some resemblance with Jacobson's derivation, and it has two steps. As the first step we 
consider massless matter fields in thermal equilibrium. Using Eq.(1.2), together with our 
result that every spacelike two-surface possesses an entropy which is one-half of its area, 
we find that massless matter fields and spacetime must obey Einstein's field equation with a 
vanishing cosmological constant. A sligthly different derivation is needed when the matter 
fields are massive, and that derivation yields Einstein's field equation with an 
unspecified cosmological constant. When we put these two derivations in together, we get
Einstein's field equation with a vanishing cosmological constant.

     Section 8 is dedicated to the high temperature limit of our model. Among other things,
one observes that in the high temperature limit the entropy of a spacelike two-surface 
depends {\it logarithmically} on its area, instead of being proportional to the area. This 
yields radical changes to the Unruh and the Hawking effects. However, it turns out, most 
curiously, that Einstein's field equation remains unchanged, no matter in which way entropy 
depends on area. This result resembles Nielsen's famous proposal of Random Dynamics, 
which states, in broad terms, that no matter what we assume about the properties of 
spacetime and matter at the Planck length scale, the low energy effects will always be 
the same \cite{kymmenen}. 

     As a starting point of our model we simply {\it assumed} that spacetime is made of 
microscopic quantum black holes with an equal spacing in their horizon area spectra. Our
treatment would not be complete without an explicit quantum mechanical model of such a 
black hole. In Section 9 we give a brief review of this kind of a model, which has earlier 
been published elsewhere \cite{yksitoista}. Beginning from the first principles of quantum mechanics  we 
get a result that the mass spectrum of a Schwarzschild black hole is discrete and bounded 
from below, and its horizon area has, in effect, an equal spacing in its spectrum.

   We close our dicussion with some concluding remarks in Section 10.

\maketitle

\section{Microscopic Quantum Black Holes}

   Black holes are extremely simple objects: After a black hole has settled down, it has 
just three classical degrees of freedom, which may be taken to be the mass, the angular momentum, 
and the electric charge of the hole \cite{kaksitoista}. Because of their simplicity, microscopic black holes are 
ideal candidates for the fundamental constituents, or atoms, of spacetime. The values taken
by just three quantities are enough to specify all of the properties of an individiual 
microscopic black hole, and in terms of these properties one may attempt to explain all of
the properties of spacetime.

   It is obvious that microscopic black holes should obey the rules of quantum mechanics.
When constructing a quantum mechanical model of spacetime out of microscopic black holes, 
the first task is therefore to find the spectra of the three classical degrees of freedom,
or obsevables, of the hole \cite{kolmetoista}. In what follows, we shall simplify the problem further, and we 
shall assume that each microscopic black hole acting as an atom of spacetime has just {\it
one} classical degree of freedom, which may be taken to be the mass of the hole. In other 
words, we shall assume that spacetime is made of microscopic, non-rotating, electrically 
neutral black holes. This means that in our model spacetime is made of microscopic 
Schwarzschild black holes.

      There have been numerous attempts to quantize the mass of the Schwarzschild black 
hole. Quite a few of them have reproduced a result, which is known as {\it Bekenstein's
proposal}. According to this proposal, which was expressed by Bekenstein already in 1974,
the event horizon area of a black hole has an equal spacing in its spectrum 
\cite{yhdeksan}. More 
precisely, Bekenstein's proposal states that the possible eigenvalues  of the event 
horizon area $A$ of a black hole are of the form:
\begin{equation}
A_n = n\gamma\ell_{Pl}^2,
\end{equation}
where $n$ is a non-negative integer, $\gamma$ is a numerical constant of order one, and 
$\ell_{Pl}$ is the Planck length of Eq.(1.1) \cite{neljatoista}.

       There are indeed very good reasons to believe in Bekenstein's proposal. First of all,
we have dimensional arguments: When a system is in a bound state, the eigenvalues of any 
quantity tend to be quantized in such a way that when we write the natural unit for the 
quantity under consideration in terms of the natural constants relevant for the system, 
then (at least in the semiclassical limit), $\hbar$ is multiplied by an integer in the 
spectrum. For Schwarzschild black holes the only relevant natural constants are $\hbar$, 
$G$ and $c$, and we find that $\ell^2_{Pl}$ is the natural unit of horizon area. Since 
$\ell^2_{Pl}$ is proportional to $\hbar$, one expects that $\ell_{Pl}^2$ is multiplied by
an integer in the spectrum. In other words, one expects that the horizon area spectrum is 
of the form given in Eq.(2.1).

      Another argument often used to support Bekenstein's proposal is Bekenstein's 
observation that horizon area is an {\it adiabatic invariant} of a black hole \cite{yhdeksan}.
 Loosely speaking, this means that the event horizon area remains invariant under very slow 
external perturbations of the hole. Adiabatic invariants of a system, in turn, are 
given by the {\it action variables} $J$ of the system. At least in the semiclassical limit
the eigenvalues of the action variables of any system are of the form:
\begin{equation}
J_n = 2\pi n\hbar,
\end{equation}
where $n$ is an integer. Because the event horizon area is an adiabatic invariant of a black
hole, it is possible that the event horizon area is proportional to one of the action 
variables of the hole. If this is indeed the case, one arrives, again, at the spectrum 
given by Eq.(2.1).

    In this paper we take Bekenstein's proposal most seriously. We shall {\it assume} that
the possible eigenvalues of the event horizon areas of the microscopic Schwarzschild 
black holes constituting spacetime are of the form:
\begin{equation}
A_n = (n + \frac{1}{2})32\pi\ell^2_{Pl},
\end{equation}
where $n=0,1,2,...$. As we shall see in Section 9, this kind of a spectrum is not merely 
idle speculation, but it is really possible to construct, using the standard rules of 
quantum mechanics, a plausible quantum mechanical model of the Schwarzschild black hole, 
which produces the area spectrum of Eq.(2.3). At this point we shall not need the details 
of that model, but the horizon area spectrum of Eq.(2.3) is sufficient for our 
considerations.

    Consider now Eq.(2.3) in more details. First of all, it is instructive to compare 
Eq.(2.3) with the energy spectrum of the field quanta in ordinary quantum field theories.
In ordinary quantum field theories in flat spacetime the standard rules of quantum 
mechanics imply the result that matter field may be thought to be constructed of 
particles, or field quanta, such that the possible energy eigenvalues of particles with
momentum $\vec k$ are of the form:
\begin{equation}
E_{n,\vec k} = (n + \frac{1}{2})\hbar\omega_{\vec k},
\end{equation}
where $n=0,1,2,...$, and $\omega_{\vec k}$ is the angular frequency corresponding to the 
momentum $\vec k$. As one may observe, there is an interesting similarity between the 
Eqs.(2.3) and(2.4): The event horizon area is quantized in exactly the same way as is the 
energy of the particles of the field. In other words, event horizon area takes the place of 
energy in our model, and we have just replaced $\hbar\omega_{\vec k}$ by $32\pi\ell_{Pl}^2$.
Indeed, one is strongly tempted to speculate that, in the fundamental level, the "quanta" 
of the gravitational field are actually microscopic quantum black holes, and that in quantum
gravity area plays a role similar to that of energy in ordinary quantum field theories.

     Since the Schwarzschild mass $M$ is the only degree of freedom of the classical 
Schwarzschild black hole, one may be interested in the mass spectrum of the microscopic 
quantum black holes as well. Because the horizon area $A$ of the Schwarzschild black hole 
is related to its mass such that:
\begin{equation}
A = \frac{16\pi G^2}{c^4}M^2,
\end{equation}
it follows from Eq.(2.3) that the eigenvalues of $M$ are of the form:
\begin{equation}
M_n = \sqrt{2n + 1}\,M_{Pl},
\end{equation}
where
\begin{equation}
M_{Pl} := \sqrt{\frac{\hbar c}{G}}\approx 2.2\times 10^{-8} kg
\end{equation}
is the Planck mass. However, as we shall see in Section 4, the Schwarzschild mass of a 
microscopic quantum black hole does not, when the hole is considered as a constituent of 
spacetime, have any sensible physical interpretation.

      In this paper we shall assume that microscopic black holes obey the standard rules of 
quantum mechanics. One of the consequences of this assumption is that the physical states 
of an individual black hole constitute a Hilbert space, which we shall denote by 
$\cal{H}_{BH}$. In this Hilbert space operates a self-adjoint event horizon area operator 
$\hat A$ with a spectrum given by Eq.(2.3), and $\cal{H}_{BH}$ is spanned by the normalized
eigenvectors $\vert \psi_n\rangle$ ($n=0,1,2,..$) of $\hat A$. In other words, an arbitrary
element $\vert \psi\rangle$ of $\cal{H}_{BH}$ may be written in the form:
\begin{equation}
\vert \psi\rangle = \sum_{n=0}^\infty c_n\vert \psi_n\rangle,
\end{equation}
where $c_n$'s are complex numbers. 

\maketitle

\section{A Model of Spacetime}

     Our next task is to construct a mathematically precise model of spacetime made of black
holes. This is a very challenging task, because when constructing such a model, we must 
think carefully, what kind of properties do we expect spacetime to possess on the fundamental
level. For instance, it is possible that at the Planck length scale spacetime does not 
have any metric properties, but the metric properties arise as statistical properties of 
spacetime at length scales very much larger than the Planck length scale. Hence, it appears 
that we must abandon, at the Planck length scale, the metric structure, and also some other 
structures familiar from classical general relativity. In our model we shall abandon the 
differentiable structure, and even the manifold structure.

   If we abandon even the manifold structure, then what are we left with? Just a set of 
points. To abandon even the last shreds of hopes of being able to construct a manifold 
structure for spacetime we shall assume that this set of points is countable. No 
manifold structure may be constructed for a countable set, because no countable set is
homeomorphic to $\Re^n$ for any $n$.

   If  we have just a countable set of points, then what can we do with those points? Well,
we may arrange them in pairs, for instance. This idea brings us to {\it graph theory} \cite{viisitoista}. 
Mathematically, a {\it graph} $G$ is defined as an ordered triple $({\cal V}, E, f)$, where 
${\cal V}$ is the set of {\it vertices} of the graph, $E$ is the set of its {\it edges}, and
$f$ is a map from $E$ to the set of unordered pairs of ${\cal V}$. In other words, for each
edge there is an unordered pair of the elements of ${\cal V}$, and therefore an edge may be 
understood as an unordered pair $\lbrace u,v\rbrace$ of the vertices of the graph. An edge 
with vertices $u$ and $v$ is denoted by $uv$. The set ${\cal V}$ of the vertices of the 
specified graph $G$ is sometimes denoted by ${\cal V}(G)$, and the set $E$ of its edges by
$E(G)$. The sets ${\cal V}(G)$ and $E(G)$ are assumed to be finite. A {\it subgraph} $G'$ 
of the given graph $G$ is a graph whose vertex and edge sets ${\cal V}(G')$ and $E(G')$
are subsets of those of the graph $G$.

   Graph theory is an important branch of mathematics which, probably because of an extreme
simplicity of its concepts and ideas, has important applications even outside pure 
mathematics. The most important practical applications of graph theory lie in network 
analysis. Graph theory also has some unsolved problems of practical value, such as the so 
called Travelling Salesman problem.

   After introducing the basic concepts of graph theory we are now prepared to express, in 
our model, a mathematically precise definition of spacetime:

   Spacetime is a graph $G$ equipped with a map 
$\varphi:{\cal V}(G)\longrightarrow {\cal H}_{BH}$. 
For every vertex $v\in {\cal V}(G)$ there exists a subgraph $G_v$ of $G$ such 
that $v\in {\cal V}(G_v)$, and the vertices $v_{k_0k_1k_2k_3}\in {\cal V}(G_v)$, where
the indices $k_\mu$ are integers for every $\mu=0,1,2,3$, have the
following properties:
\begin{subequations}
\begin{eqnarray}
v_{k_0k_1k_2k_3}v_{(k_0+1)k_1k_2k_3} &\in& E(G_v) \forall k_0\in J_{k_1k_2k_3},\\
v_{k_0k_1k_2k_3}v_{k_0(k_1+1)k_2k_3} &\in& E(G_v) \forall k_1\in J_{k_0k_2k_3},\\
v_{k_0k_1k_2k_3}v_{k_0k_1(k_2+1)k_3} &\in& E(G_v) \forall k_2\in J_{k_0k_1k_3},\\
v_{k_0k_1k_2k_3}v_{k_0k_1k_2(k_3+1)} &\in& E(G_v) \forall k_3\in J_{k_0k_1k_2},\\
v_{k_0k_1k_2k_3} = v_{k'_0k'_1k'_2k'_3} &\iff& k_\mu = k'_\mu \forall \mu=0,1,2,3.
\end{eqnarray}
\end{subequations}
In these equations we have defined:
\begin{subequations}
\begin{eqnarray}
J_{k_1k_2k_3} &:=& \lbrace N_{k_1k_2k_3}, N_{k_1k_2k_3}+1,...,M_{k_1k_2k_3}-1\rbrace,\\
J_{k_0k_2k_3} &:=& \lbrace N_{k_0k_2k_3}, N_{k_0k_2k_3}+1,...,M_{k_0k_2k_3}-1\rbrace,\\
J_{k_0k_1k_3} &:=& \lbrace N_{k_0k_1k_3},N_{k_0k_1k_3}+1,...,M_{k_0k_1k_3}-1\rbrace,\\
J_{k_0k_1k_2} &:=& \lbrace N_{k_0k_1k_2},N_{k_0k_1k_2}+1,...,M_{k_0k_1k_2}-1\rbrace,
\end{eqnarray}
\end{subequations}
where $N_{k_0k_1k_2}$, $M_{k_0k_1k_2}$, etc. are non-negative integers such that for all 
$\alpha,\beta,\gamma\in\lbrace 0,1,2,3\rbrace$, where $\alpha<\beta<\gamma$, we have 
$M_{k_\alpha k_\beta k_\gamma}>N_{k_\alpha k_\beta k_\gamma}$.

    Consider now our definition of spacetime in details. First of all, why should spacetime
be a graph? The answer to this question lies in the simplicity of the concept of graph. 
Indeed, a graph is just a set of points equipped with a rule which tells, whether the two given 
points belong to the same edge or not. In other words, there operates a 
{\it binary relation}, the simplest possible non-trivial relation, in the set of the 
unordered pairs of the elements of ${\cal V}(G)$. So we have already in this early stage 
of our project achieved a possible realization of Wheeler's famous "it-from-bit"-proposal
which states that the fundamental theory of nature should be based on binary relations 
\cite{kuusitoista}.

    It is an important feature of Einstein's general theory of relativity that it is a 
{\it local} theory of spacetime in the sense that nothing is assumed about the global 
properties, such as the topology, of spacetime. Instead, certain local properties are 
assumed. For instance, one assumes that every point $P$ of spacetime has a neighborhood 
$U(P)$, which is homeomorphic to $\Re^4$. To preserve this local nature of classical 
general relativity in our model of spacetime, we have focussed our attention to the 
properties of the subgraphs $G_v$ of the graph $G$, instead of considering the graph $G$
itself. The subgraph $G_v$ having the vertex $v$ as one of its vertices may be viewed as 
a graph theoretic counterpart of the topological concept of neighborhood.

   Our next observation is that every vertex of the subgraph $G_v$ has been identified by
{\it four integers} $k_\mu$, where $\mu=0,1,2,3$. This is an attempt to bring something 
resembling the manifold structure of classical spacetime into our model: In classical 
spacetime every point has a neighborhood, where each point may be identified by four reals
$x^\mu$ $(\mu=0,1,2,3)$, whereas in our model for every vertex $v$ there exists a subgraph
$G_v$ such that the vertices of this subgraph may be identified by four integers. For fixed
$k_1$, $k_2$ and $k_3$, the integers $N_{k_1k_2k_3}$ and $M_{k_1k_2k_3}$ set the lower-, and 
the upper bounds, respectively, for the allowed values of the integer $k_0$. Accordingly,
the integers $N_{k_0k_2k_3}$ and $M_{k_0k_2k_3}$ set the lower-, and the upper bounds for 
the integer $k_1$ for fixed $k_0$, $k_2$ and $k_3$, and so on.

   Let us now turn our attention to Eq.(3.1a). That equation states that for all fixed 
$k_1$, $k_2$ and $k_3$, $v_{k_0k_1k_2k_3}v_{(k_0+1)k_1k_2k_3}$ is one of the edges of the
subgraph $G_v$ for all $k_0=N_{k_1k_2k_3},N_{k_1k_2k_3}+1,...,M_{k_1k_2k_3}-1$. In other 
words, we have an alternating sequence of vertices and edges, beginning and ending with a 
vertex, in which each vertex is incident to the two edges that precede and follow it in the 
sequence, and the vertices that preceed and follow an edge are end vertices of the edge. 
In graph theory, such a sequence of vertices and edges is known as a {\it walk}. It follows
from Eq.(3.1e) that the first and the last vertices of the walk are different, and every
vertex of the walk is incident to at most two edges of the walk. Such a walk is known as 
a {\it path}. So we find that for fixed $k_1$, $k_2$ and $k_3$ we have a certain path in
the subgraph $G_v$. The vertices of this path are identified by the integer $k_0$. 
Accordingly, for fixed $k_0$, $k_2$ and $k_3$ we have a path in $G_v$, whose vertices are
identified by the values of the integer $k_1$, and so on. These kind of paths may be 
viewed as graph theoretic analogues of {\it coordinate curves} in differentiable manifolds.
We shall introduce a terminology, where we say that an edge of the subgraph $G_v$ is 
{\it timelike}, if the values taken by $k_{1,2,3}$ are the same at its endvertices and the 
values taken by $k_0$ are different, and the edge is {\it spacelike}, if the values taken 
by $k_0$ at its end vertices are the same. The subgraph $\sigma_{k_0}$ of $G_v$, where 
$k_0=constant$ for all of the vertices of $\Sigma_{k_0}$, we shall call a {\it spacelike 
hypergraph} of $G_v$. It should be noted that at this stage these are just names; they have
nothing to do with space and time themselves. Later, in Section 5, we shall introduce the 
concept of time into our model as a macroscopic quantity which, however, has no relevance
at the Planck length scale.

    Finally, we have in spacetime a map $\varphi$ from the vertices of the graph $G$ to the
Hilbert space ${\cal H}_{BH}$ of the black hole states $\vert\psi\rangle$. This means that
we associate with every vertex $v$ of $G$ a unique quantum state of a microscopic quantum
black hole. If one likes, one may understand this in such a way that at every vertex of $G$
we have a microscopic quantum black hole.

\maketitle

\section{Statistics of Spacetime}

   We have now constructed a mathematically precise model of spacetime, and identified its 
fundamental constituents, the microscopic quantum black holes, together with their quantum
states. The next task is to investigate the statistics of spacetime. The fundamental 
problem of the statistical physics of any system is to identify the micro- and the 
macrostates of the system. During the recent years it has become increasingly clear that the 
gravitational properties of any three-dimensional region of space may be read off from 
its {\it two-dimensional boundary}. For instance, it has been shown by Padmanabhan et al. 
that Einstein's field equation may be obtained by varying, in an appropriate manner, the 
boundary term of the Einstein-Hilbert action \cite{seitsemantoista}. So it appears that 
{\it spacelike two-surfaces} play a fundamental role in gravitational physics.

    Since spacetime in our model is assumed to be a graph, the first question is: What is 
the counterpart of a spacelike two-surface in our model? Intuitively, it is fairly obvious 
that those subgraphs of the graph $G_v$ presented in the previous Section, where the index 
$k_0$, and one of the indecies $k_{1,2,3}$ labelling the vertices $v_{k_0k_1k_2k_3}$ of the 
graph $G_v$ are constants, are examples of such counterparts: Every vertex of those 
subgraphs is specified by exactly two free indices in the same way as every point on a 
two-surface is specified by two coordinates. In other words, points on a two-surface are 
replaced by vertices, and the coordinates of the points are replaced by the indices of the 
vertices. So we find that the counterpart of a spacelike two-surface in our model should 
possess properties similar to the graph ${\cal U}_2$, which we shall define as a graph with
vertices $u_{k_1k_2}$ such that
\begin{subequations}
\begin{eqnarray}
u_{k_1k_2}u_{(k_1+1)k_2} &\in& E({\cal U}_2) \forall k_1\in J_{k_2},\\
u_{k_1k_2}u_{k_1(k_2+1)} &\in& E({\cal U}_2) \forall k_2\in J_{k_1},\\
u_{k_1k_2} = u_{k'_1k'_2} &\iff& k_\mu = k'_\mu \forall \mu=1,2,
\end{eqnarray}
\end{subequations}
where we have defined:
\begin{subequations}
\begin{eqnarray}
J_{k_1} &:= \lbrace N_{k_1}, N_{k_1}+1,...,M_{k_1}-1\rbrace,\\
J_{k_2} &:= \lbrace N_{k_2}, N_{k_2}+1,...,M_{k_2}-1\rbrace,
\end{eqnarray}
\end{subequations}
where $N_{k_{1,2}}$ and $M_{k_{1,2}}$ are non-negative integers such that $M_{k_1}>N_{k_1}$, and 
$M_{k_2}>N_{k_2}$. Indeed, the vertices of 
the graph are identified by exactly two indices, which may take certain integer values only.

   In what follows, we shall adopt, for the sake of brevity and simplicity, the name {\it
spacelike two-graph} for the counterpart of spacelike two-surface in our model of spacetime.
To construct a precise definition of this counterpart we first define the concept of {\it
embedding} in graph theory: If $G_1$ and $G_2$ are graphs, then an {\it embedding $G_1$ of
$G_2$} is an injection $f:{\cal V}(G_2)\longrightarrow {\cal V}(G_1)$ such that every edge 
in $E(G_2)$ corresponds to a path, disjoint from all other such paths, in $G_1$.

   We are now prepared to write the definition of a spacelike two-graph of spacetime $G$: 
Let $G_{k_0}$ be a spacelike hypergraph of $G$. In that case a spacelike two-graph of 
spacetime is an embedding $G_{k_0}$ of ${\cal U}_2$. As one may observe, our definition of 
a spacelike two-graph is very closely related to the definition of a spacelike two-surface 
of classical spacetime: In classical spacetime a spacelike two-surface is an 
embedding of an open subset $U$ of $\Re^2$ to a spacelike hypersurface of spacetime,
whereas in our definition we have replaced the set $U$ by the set ${\cal U}_2$, and 
spacelike hypersurface by a spacelike hypergraph. In what follows, we shall denote spacelike
two-graphs of spacetime by $\Sigma^{(2)}$. 

  We shall now enter the statistics of $\Sigma^{(2)}$. The first task is to find a 
thermodynamical, macroscopic variable which would label the macroscopic states of 
$\Sigma^{(2)}$. In ordinary statistical physics the most important macroscopic quantity
used for the labelling of the macrostates of the system is the {\it energy} of the system.
However, as we saw in Section 2, there are good grounds to believe that in quantum gravity
{\it area} takes the place of energy as the fundamental quantity. We shall therefore label 
the macroscopic states of $\Sigma^{(2)}$ by means of its area $A$ (the precise definition of 
this concept becomes later): When the area $A$ of $\Sigma^{(2)}$ changes, so does the 
macroscopic state of $\Sigma^{(2)}$. 

   After solving the problem of the macrostates of $\Sigma^{(2)}$, we shall now enter to the 
problem of its microstates. We shall assume that the microscopic quantum black holes 
residing at the vertices of $\Sigma^{(2)}$ will satisfy the following {\it independence
postulates}:

(IP1): The quantum states of the microscopic quantum black holes associated with the 
vertices of $\Sigma^{(2)}$ are independent of each other.

(IP2): The ground states of the microscopic quantum black holes do not contribute to the 
total area of $\Sigma^{(2)}$. 

(IP3): When a hole is in the $n$'th excited state, it contributes to $\Sigma^{(2)}$ an
area, which is directly proportional to $n$. 

(IP4): The total area $A$ of $\Sigma^{(2)}$ is the sum of the areas contributed by the
 black holes associated with the vertices of $\Sigma^{(2)}$, to the area of 
$\Sigma^{(2)}$.

    No doubt, it is rather daring to assume that the black holes on $\Sigma^{(2)}$ are 
independent of each other. At this stage of research, however, that is an absolutely 
necessary assumption to make progress. Actually, the postulate (IP1) is very similar
to the assumption of an asymptotic freedom of quarks in QCD: When quarks come very close
to each other, they may be considered, essentially, as free and independent particles.
As such the postulate (IP1) is not necessarily entirely unphysical. Neither is the postulate
(IP2): In ordinary quantum field theories the vacuum states of the particles of the field 
do not contribute to the total energy of the field. Since in our model area takes the place
of energy, it is only natural to assume that the vacuum states of the microscopic quantum
black holes do not contribute to the area of $\Sigma^{(2)}$. When put in together, the 
postulates (IP1)-(IP4) imply that the possible eigenvalues of the total area $A$ of 
$\Sigma^{(2)}$ are of the form:
\begin{equation}
A = \alpha(n_1 + n_2 + ... + n_N),
\end{equation}
where the coefficient $\alpha$ is a pure number to be determined later, and the quantum 
numbers $n_1$, $n_2$,...,$n_N$ label the quantum states of the microscopic quantum black 
holes lying on $\Sigma^{(2)}$. 

   We are now prepared to enter the statistics of microscopic quantum black holes on 
$\Sigma^{(2)}$. We shall assume that the holes obey the following 
{\it statistical postulates}: 

(SP1): The microstates of $\Sigma^{(2)}$ are uniquely determined by the combinations of 
the non-vacuum horizon area eigenstates of the microscopic quantum black holes on 
$\Sigma^{(2)}$. 

(SP2): Each microstate of $\Sigma^{(2)}$ yielding the same area eigenvalue of $\Sigma^{(2)}$
corresponds to the same macrostate of $\Sigma^{(2)}$. 

 Using these postulates one may calculate the {\it statistical weight} $\Omega$ of a given
macrostate of $\Sigma^{(2)}$, or the number of microstates corresponding to that macrostate.
It follows from Eq.(4.3) that $\Omega$ is the number of ways to express $\frac{A}{\alpha}$ 
as a sum of positive integers $n_1$, $n_2$,...,$n_N$, where $N$ is the number of microscopic 
quantum black holes on $\Sigma^{(2)}$. More precisely, $\Omega$ is the number of strings
$(n_1,n_2,...,n_m)$, where $1\leq m\leq N$ and $n_1,...,n_m\in \lbrace 1,2,3,...\rbrace$ 
such that:
\begin{equation}
n_1 + n_2 +...+ n_m = \frac{A}{\alpha}.
\end{equation}

    Now, what is $\Omega$? Since $N$ is fixed, this is a fairly non-trivial problem, because
the number of the positive integers $n_i$ $(n_i=1,2,...,m)$ is bounded by $N$. However, if 
we assume that 
\begin{equation}
N>>\frac{A}{\alpha},
\end{equation}
the limitations posed by the upper-boundedness of $m$ by $N$ are removed. Physically, the 
condition (4.5) means that {\it most microscopic quantum black holes on $\Sigma^{(2)}$ are 
in the vacuum}. In this paper we shall call the limit (4.5) as the {\it low temperature 
limit}. It should be noted that we have not defined any concept of spacetime temperature 
yet. We shall adopt the name "low temperature limit" simply because physical systems possess
a general feature that in low temperatures most constituents of a system are at as low 
states as possible. In this Section we shall consider the low temperature limit only. In 
Section 8 we shall consider the general case, where the condition (4.5) is removed. 

    It is easy to calculate $\Omega$ in the low temperature limit. In this limit $\Omega$ is
simply the number of ways to write $\frac{A}{\alpha}$ as a sum of an unlimited number of
positive integers. It is easy to show that there are $2^{A/{\alpha}-1}$ ways to do that. In 
other words, we have 
\begin{equation}
\Omega = 2^{\frac{A}{\alpha} -1}
\end{equation}
in the low temperature limit \cite{kahdeksantoista}. This result has an attractive feature from an information 
theoretic point of view: If a system has $2^n$ possible states, it may accomodate $n$ bits
of information \cite{yhdeksantoista}. Hence it follows that, in the low temperature limit, the number of bits of
information $\Sigma^{(2)}$ is able to accomodate is
\begin{equation}
H = \frac{A}{\alpha}-1,
\end{equation}
and because $\frac{A}{\alpha}$ is assumed to be very big, we may write, as an excellent 
approximation:
\begin{equation}
H = \frac{A}{\alpha}.
\end{equation}
Since $\frac{A}{\alpha}$ gives the number of quanta of area of $\Sigma^{(2)}$, we may 
therefore conclude that in the low temperature limit {\it each quantum of area carries
exactly one bit of information}. This is a very interesting result, because it reminds us,
again, of Wheeler's famous "it-from-bit"-proposal. Here we have seen that this proposal may 
indeed be realized: When most constituents of $\Sigma^{(2)}$ are in the ground state, 
$\Sigma^{(2)}$ may accomodate an integer number of bits of information. 

  Using Eq.(4.3) we may calculate the {\it entropy} of a given macroscopic state of 
$\Sigma^{(2)}$. We find that, in the low temperature limit, the entropy is:
\begin{equation}
S = \ln\Omega = \frac{\ln 2}{\alpha}A,
\end{equation}
where we have, again, assumed that the condition (4.5) holds. As one may observe, entropy is
proportional to area. This is a very satisfying result, because we know from black hole
physics that the entropy of a black hole is proportional to its event horizon area.

   In this Section we have considered area and entropy of {\it spacelike} two-graphs only.
Those two-graphs, in turn, correspond in a certain way to the spacelike two-surfaces of 
classical spacetime. What can we say about the areas of those two-graphs, which are {\it not}
spacelike? In particular, is the relationship between the area of a general two-graph
(by which we mean an embedding of the graph ${\cal U}_2$ into {\it any} subgraph,
satisfying the postulates of Eqs.(3.1) and (3.2), of the 
spacetime $G$), and the quantum states of the microscopic quantum black holes lying on its 
vertices similar to the one expressed in Eq.(4.3) even when that two-graph is not spacelike?

   There are good grounds to believe that this is indeed the case: For instance, one may 
expect that the life time of a microscopic quantum black hole is proportional to its mass
parameter $M$. More precisely, the proper time measured by an observer in a radial free fall
from the past to the future singularity through the bifurcation two-sphere in Kruskal 
spacetime is $2\pi M$ (see Section 9). Since the diameter of a Schwarzschild black hole is
$4M$, one may expect that each microscopic black hole occupies an area proportional to 
$M^2$ even on timelike two-surfaces of spacetime. Moreover, since the eigenvalues of $M$
are those given in Eq.(2.6), one may expect that the relationship between the area $A$ of 
{\it any} two-graph and the quantum states of the microscopic quantum black holes lying at 
its vertices is indeed given by Eq.(4.3). We accept this assumption as one of the postulates of our 
model, and see, where it will take us.

 \maketitle

\section{Emergence of Classical from Quantum Spacetime}

   Whatever the correct quantum mechanical model of spacetime may be, it should ultimately
reproduce Einstein's field equation in an appropriate limit. Einstein's field equation,
however, is written in {\it classical} spacetime, and therefore the question is: How does 
classical spacetime emerge from quantum spacetime? Presumably the relationship between 
classical and quantum spacetime is analogous to that of a macroscopic body, and its 
atomic substructure: the properties of a macroscopic body arise as sort of statistical 
averages of the properties of its atoms. In the same way, the properties of classical 
spacetime should arise as statistical averages of its microscopic constituents.

   Classical spacetime is an approximation of quantum spacetime in a long distance and,
presumably, low temperature limit. The fundamental concepts of classical spacetime, such as
metric and curvature, are absent in quantum spacetime. The object of this Section is to find
the precise relationship between the concepts used in classical and quantum spacetimes.

   Before we proceed any further, we must define some new concepts. The first concept we 
define is the {\it boundary} $\partial\Sigma^{(2)}$ of a two-graph $\Sigma^{(2)}$ of 
spacetime. To define this concept we must, in turn, first define the concept of the boundary
$\partial{\cal U}_2$ of the graph ${\cal U}_2$ we defined in Eqs.(4.1) and (4.2). We first 
define the set ${\cal W}_{\cal U}$ of the vertices $u_{k_1k_2}$ of ${\cal U}_2$ such that
for fixed $k_2$:
\begin{equation}
k_1\in \lbrace N_{k_2},M_{k_2}\rbrace,
\end{equation}
and for fixed $k_1$:
\begin{equation}
k_2 \in \lbrace N_{k_1}, M_{k_1}\rbrace,
\end{equation}
where the non-negative integers $N_{k_{1,2}}$ and $M_{k_{1,2}}$ were defined in 
Eqs.(4.1) and (4.2).
The boundary $\partial{\cal U}_2$ of the graph ${\cal U}_2$ is defined as that subgraph of 
the graph ${\cal U}_2$, which has the property 
${\cal V}(\partial{\cal U}_2) = {\cal W}_{\cal U}$. In other words, the vertices of 
$\partial{\cal U}_2$ consist of the elements of ${\cal W}_{\cal U}$. It is easy to see that
$\partial{\cal U}_2$ is a closed, simple walk through the elements of ${\cal W}_{\cal U}$.

    It is now easy to define the boundary $\partial\Sigma^{(2)}$ of the two-graph 
$\Sigma^{(2)}$ of spacetime: If $\Sigma^{(2)}$ is an embedding 
$f:{\cal U}_2\longrightarrow G$, then $\partial\Sigma^{(2)}$ is an embedding
$f:\partial{\cal U}_2 \longrightarrow G$. In other words,
\begin{equation}
\partial\Sigma^{(2)} := f\vert_{\partial{\cal U}_2}.
\end{equation}
So we find that when ${\cal U}_2$ is embedded in $G$, then the boundary of ${\cal U}_2$ is 
mapped to the boundary of $\Sigma^{(2)}$.

   The next concept we define is a {\it discrete geodesic}: A discrete geodesic between 
the vertices $v_1$ and $v_2$ of spacetime is the path $\gamma$ which joins these two
vertices with the least possible number of edges. In other words, when we are trying to find
a discrete geodesic between two given vertices, we first take all paths joining these 
vertices, and then pick up the one which has the smallest number of edges. Intuitively, one
might expect that after one has introduced the concept of length in spacetime, then the 
discrete geodesic would be the shortest path joining the two given vertices of spacetime, 
at least when the number of vertices per path is very large.

  By means of the concept of discrete geodesic we may define the concept of
 {\it discrete triangle}: Let us pick up three vertices $v_1$, $v_2$ and $v_3$ such that 
the discrete geodesics joining these vertices to each other are all
non-intersecting. In that case we get a closed, simple path, which we shall denote by
$K_{123}$. A {\it discrete triangle determined by the vertices $v_1$, $v_2$ and $v_3$} is
denoted by $\Delta_{123}$, and it is defined as the two-graph, which has $K_{123}$ as its
 boundary, and which has the smallest number of vertices. In other words, we find all those
two-graphs of $G$, which have $K_{123}$ as their common boundary, and then pick up the one,
which has the smallest number of vertices. Intuitively, one might expect that of all 
two-graphs having $K_{123}$ as their common boundary, $\Delta_{123}$ is the one, which 
has the smallest area, at least when the number of vertices is very large.

   We are now prepared to enter the project of reconstruction of classical spacetime, in an 
appropriate limit, from quantum spacetime. As the first step let us consider a possibility
of constructing a system of coordinates in spacetime. Let us pick up a vertex $v$ of 
spacetime. According to the definition of our spacetime model there exists a subgraph $G_v$
of spacetime such that $v\in{\cal V}(G_v)$, and the vertices $v_{k_0k_1k_2k_3}$ of $G_v$ 
have the properties given in Eqs.(3.1) and (3.2). Let us denote the number of the vertices 
of $G_v$ by $N$. At macroscopic length scales $N$ is enormous: For instance, if the diameter
of the spacetime region under consideration is, in spatial as well as in the temporal 
directions of the order of 1 meter, then $N$ is presumably of the order of $(10^{35})^4$, or 
$10^{140}$. Yet 1 meter is still a length scale at which the classical general relativistic
effects on spacetime geometry, in normal situations, are negligible. The only exceptions 
are the black hole and the big bang singularities. So we find that if we are trying to get
general relativity out from our quantum mechanical model of spacetime, then statistical and
thermodynamical methods, rather than the methods of microscopic physics, should apply. 

   Keeping these things in mind we now define the maps 
$\chi^\mu:{\cal V}(G_v)\longrightarrow\Re$ such that
\begin{equation}
\chi^\mu(v_{k_0k_1k_2k_3}) := \frac{k_\mu}{N^{1/4}},
\end{equation}
for every $v_{k_0k_1k_2k_3}\in {\cal V}(G_v)$, and $\mu=0,1,2,3$. The maps $\chi^\mu$ may 
be understood as the {\it coordinates} of the vertices of $G_v$. The graphs, where 
all indices $k_\mu$, except one, of the vertices $v_{k_0k_1k_2k_3}$ are constant, 
represent {\it coordinate curves} in $G_v$. If $k_\mu$ gets a change $\Delta k_\mu$, then 
the corresponding change of the coordinate $\chi^\mu$ is
\begin{equation}
\Delta \chi^\mu = \frac{\Delta k_\mu}{N^{1/4}}.
\end{equation}
As one may observe, for finite $\Delta k_\mu$ the change of $\chi^\mu$ becomes very small,
when $N$ becomes very large. In other words, the continuum limit is achieved, when $N$ goes 
to infinity. In this limit $\Delta \chi^\mu$ becomes to the ordinary differential $d\chi^\mu$. 
The number $N$ therefore keeps the record of how close we are to the continuum limit. In the
continuum limit we should recover the concepts and the results of classical general
relativity.

    Our next task would be to define concepts analogous to {\it tensors} in classical general
relativity. In classical general relativity the starting point of the definition of tensors 
is the definition of the concept of {\it tangent space}. Each point $P$ of classical 
spacetime has its own tangent space $T_P$ which approximates the neighborhood of the point 
$P$ at length scales, where the total curvature of spacetime is negligible. When 
constructing an analogue of the concept of tangent space in our spacetime model we should 
find a space which approximates a spacetime region under consideration at length scales, 
where the 
effects of both the Planck scale quantum fluctuations, and of classical general relativity 
on spacetime curvature are negligible. Our journey to the precise definition of such a 
concept is a rather long one, and it has several stages and steps.

   As the first step we pick up a real number $\lambda\in(0,1)$,
and we define the vertices $w$ and $v_\mu$ $\in{\cal V}(G_v)$ $(\mu=0,1,2,3)$ such that:
\begin{subequations}
\begin{eqnarray}
w &:=& v_{k_0k_1k_2k_3},\\
v_0 &:=& v_{(k_0+\lambda N^{1/4})k_1k_2k_3},\\
v_1 &:=& v_{k_0(k_1+\lambda N^{1/4})k_2k_3},\\
v_2 &:=& v_{k_0k_1(k_2+\lambda N^{1/4})k_3},\\
v_3 &:=& v_{k_0k_1k_2(k_3+\lambda N^{1/4})}.
\end{eqnarray}
\end{subequations}
Making $\lambda$ small we may bring the vertices $v_\mu$ as close to the vertex $w$ as we 
please. We have assumed, of course, that $\lambda N^{1/4}$ is an integer, since otherwise 
Eq.(5.6) would not make sense. We have also assumed that the discrete geodesics connecting 
every one of the vertices $w$ and $v_\mu$ to every other vertex are non-intersecting and 
they are subgraphs of $G_v$, as 
well as are the discrete triangles which result when we arrange the vertices $w$ and 
$v_\mu$ in groups which contain three vertices each.

  As the second step we define an injective map 
$\tilde{f}:\lbrace w,v_0,v_1,v_2,v_3\rbrace\longrightarrow \Re^4$ such that the five points
\begin{subequations}
\begin{eqnarray}
\tilde{w} &:= \tilde{f}(w),\\
\tilde{v}_\mu &:= \tilde{f}(v_\mu)
\end{eqnarray}
\end{subequations}
are linearly independent for all $\mu=0,1,2,3$. Moreover, we pose for the map $\tilde{f}$ an 
additional requirement that the areas of the triangles, or two-faces, of the 
{\it geometrical four-simplex}
\begin{equation}
\sigma_w :=\tilde{w}\tilde{v}_0\tilde{v}_1\tilde{v}_2\tilde{v}_3
\end{equation}
are equal to the areas of the corresponding discrete triangles in $G_v$. In other words, we
require that the area of the two-face $\tilde{w}\tilde{v}_0\tilde{v}_1$ is equal to the area
of the discrete triangle with vertices $w$, $v_0$ and $v_1$,  the area of the two-face 
$\tilde{v}_0\tilde{v}_1\tilde{v}_2$ is equal to the area of the discrete triangle with 
vertices $v_0$, $v_1$ and $v_2$, and so on. Geometrical simplices are basic objects of
{\it algebraic topology}, and the geometrical four-simplex $\sigma_w
=\tilde{w}\tilde{v}_0\tilde{v}_1\tilde{v}_2\tilde{v}_3$ is defined as a convex hull of the
points ${\tilde w}$, ${\tilde v}_0$, ${\tilde v}_1$, ${\tilde v}_2$ and ${\tilde v}_3$
 \cite{kaksikymmenta}.

    Now, our geometrical four-simplex $\sigma_w$ has
\begin{equation}
\left(\begin{array}{cc}
5\\
2\end{array}\right) = 10
\end{equation}
one-faces or {\it edges}, and
\begin{equation}
\left(\begin{array}{cc}
5\\
3\end{array}\right) = 10
\end{equation}
triangles. As one may observe, the number of triangles in $\sigma_w$ is the same as is the 
number of edges, and there is a one-to-one correspondence between the triangles and the 
edges. This is a phenomenon which happens in four dimensions only. It does not happen in 
any other number of dimensions.

   Because of the one-to-one correspondence between the triangles and the edges of the 
four-simplex $\sigma_w$, one might expect that the lengths of the edges of $\sigma_w$
could be expressed in terms of the areas of its triangles. This is indeed the case, but 
before we are able to write down the precise relationhips between edge lengths and triangle
areas, we must construct a metric inside the four-simplex $\sigma_w$.

   The four-simplex $\sigma_w$ is a four-dimensional subset of $\Re^4$, and therefore it is
possible to choose in $\sigma_w$ a {\it flat} four-dimensional system of coordinates, which 
covers the whole $\sigma_w$. This system of coordinates may be chosen to be either the 
ordinary Cartesian system of coordinates with a Euclidean signature in the metric, or the
flat Minkowski system of coordinates. In what follows, we shall consider both of these 
options.

   We begin with the system of coordinates, which produces the Euclidean signature in the 
metric. In that case there is no difference between space and time. We shall denote the 
coordinates by $X^I$ $(I=0,1,2,3)$, and the corresponding base vectors we shall denote by 
${\hat e}_I$. Since spacetime is assumed to be flat and Euclidean, the inner product between
the vectors ${\hat e}_I$ obeys the rule:
\begin{equation}
{\hat e}_I\bullet{\hat e}_J = \delta_{IJ},
\end{equation}
for all $I,J=0,1,2,3$. In other words, the base vectors ${\hat e}_I$ are orthonormal. The
inner product of an arbitrary vector 
\begin{equation}
{\vec V} := V^I{\hat e}_I
\end{equation}
(Einstein's sum rule has been used) with itself is
\begin{equation}
V^2 := {\vec V}\bullet{\vec V} = V^I V^J\delta_{IJ} = (V^0)^2 + (V^1)^2 + (V^2)^2 + (V^3)^2.
\end{equation}
We fix our system of coordinates such that its origin lies at the vertex ${\tilde w}$, and
\begin{subequations}
\begin{eqnarray}
X^0(P) &=& 0\,\,\forall P\in{\tilde w}{\tilde v}_1{\tilde v}_2{\tilde v}_3,\\
X^1(P) = X^2(P) &=& 0\,\,\forall P\in{\tilde w}{\tilde v}_1{\tilde v}_2,\\
X^1(P) = X^2(P) = X^3(P) &=& 0\,\,\forall P\in{\tilde w}{\tilde v}_1.
\end{eqnarray}
\end{subequations}
This means that the vector ${\hat e}_1$ is parallel to the edge ${\tilde w}{\tilde v}_1$, 
which joins the vertices ${\tilde w}$ and ${\tilde v}_1$, the vector ${\hat e}_2$ lies on
the two-face, or triangle ${\tilde w}{\tilde v}_1{\tilde v}_2$, being ortohogonal to the 
edge ${\tilde w}{\tilde v}_1$ and the vector ${\hat e}_3$ lies on the three-face, or 
tetrahedron ${\tilde w}{\tilde v}_1{\tilde v}_2{\tilde v}_3$ such that ${\hat e}_3$ is 
perpendicular to the triangle ${\tilde w}{\tilde v}_1{\tilde v}_2$. Finally, the vector
${\hat e}_0$ is orthogonal to the three-space determined by the tetrahedron 
${\tilde w}{\tilde v}_1{\tilde v}_2{\tilde v}_3$.

  We shall now introduce a notation, where
\begin{equation}
\tilde{v}_4 := \tilde{w},
\end{equation}
and
\begin{equation}
s_{ab} := \vec{v}_{ab}\bullet\vec{v}_{ab},
\end{equation}
where $\vec{v}_{ab}$ is the vector which joins the vertex $\tilde{v}_a$ to the vertex 
$\tilde{v}_b$ for all $a,b=0,1,2,3,4$. It is easy to show that in this notation between 
the area $A_{abc}$ of a triangle with vertices $\tilde{v}_a$, $\tilde{v}_b$ and 
$\tilde{v}_c$, and the quantities $s_{ab}$ defined in Eq.(5.16) there is a relationship:
\begin{equation}
 4s_{ac}s_{bc} + (s_{ac} + s_{bc} - s_{ab})^2 = 16A_{abc}^2.
\end{equation}
As a whole, this relationship consists of a system of 10 equations and 10 unknowns. It is 
therefore possible to solve the quantities $s_{ab}$ from these equations 
in terms of the areas $A_{abc}$. The areas $A_{abc}$, in turn, are equal to the areas of the
corresponding discrete triangles in $G_v$. Since the areas of the discrete triangles depend
on the quantum states of the microscopic black holes lying at their vertices as in Eq.(4.3),
we find that the lengths of the edges of the four-simplex $\sigma_w$ may be reduced
to the quantum states of the microscopic black holes.

  It is also possible to write Eq.(5.17) in terms of the coordinates of the vertices 
${\tilde v}_a$ ($a=0,1,2,3,4$) of the four-simplex $\sigma_w$. If we denote the coordinates 
$X^I$ of the vertex ${\tilde v}_a$ in a system of coordinates defined in Eqs.(5.11)-(5.14)
by $X^I_a$, we may write:
\begin{equation}
s_{ab} = \sum_{I=0}^3 (X^I_a - X^I_b)^2,
\end{equation}
for every $a,b=0,1,2,3,4$. If we substitute this expression for $s_{ab}$ in Eq.(5.17), we 
get a system of 10 equations for 20 unknowns $X^I_a$. However, it follows from Eq.(5.14) 
that we have:
\begin{subequations}
\begin{eqnarray}
X^I_4 &=& 0\,\,\forall I=0,1,2,3,\\
X^0_2 = X^2_1 = X^3_1 &=& 0,\\
X^0_2 = X^3_2 &=& 0,\\
X^0_3 &=& 0.
\end{eqnarray}
\end{subequations}
As a whole we therefore have only 10 unknowns to solve from 10 equations. In other words, if
one substitutes Eq.(5.18) in Eq.(5.17), one may solve the coordinates $X^I_a$ of the 
vertices ${\tilde v}_a$ of the four-simplex $\sigma_w$ in terms of the areas of its 
two-faces.

  So far we have assumed everywhere that the metric inside the four-simplex $\sigma_w$ is 
flat and Euclidean. Using this assumption we found that it is possible to reduce the lengths 
of the edges to the area eigenvalues of the microscopic black holes. In other words it seems 
that, at least at macroscopic length scales, the concept of length may be reduced back to 
the quantum states of the microscopic black holes, which constitute spacetime. However, 
since the four-simplex $\sigma_w$ approximates, in certain very specific sense which will be
explained in details in a moment, spacetime at macroscopic length scales, one might expect
that instead of having a flat Euclidean metric, we should have a flat Minkowski metric
inside the four-simplex $\sigma_w$. In other words, one expects that the inner product 
between the vectors ${\hat e}_a$ should obey the rule
\begin{equation}
{\hat e}_I\bullet{\hat e}_J = \eta_{IJ},
\end{equation}
where $\eta_{IJ}:=diag(-1,1,1,1)$, instead of the rule of Eq.(5.11). One might also feel 
that this rule should be followed right from the beginning. More precisely, it would appear 
natural to write Eqs.(5.17) and (5.18) using the Minkowski, instead of the Euclidean metric. 

  Although this kind of an approach may seem attractive at first, it has problems of its 
own: In spacetime equipped with a flat Minkowski metric the area of an arbitrary lightlike
two-surface is zero, and therefore one expects that all microscopic quantum black holes 
on that two-surface should be in the ground state. However, every microscopic black hole of
spacetime belongs to some lightlike two-surface, and therefore it follows that every hole
of spacetime should be in a ground state, which is certainly not true.

   Another approach to the problem of the signature of the metric should therefore be 
employed. In this paper we shall adopt an approach, where we take first a Euclidean metric 
inside the four-simplex $\sigma_w$. We solve the quantities $s_{ab}$ from Eq.(5.17) and,
moreover, the coordinates $X^I_a$ of the vertices of $\sigma_w$ by means of Eq.(5.18). After
finding the coordinates of the vertices we perform in $\sigma_w$ the {\it Wick rotation}, 
where we replace the real coordinate $X^0_a$, which represents time in $\sigma_w$, by an
{\it imaginary} coordinate $iX^0_a$. Mathematically, the Wick rotation is equivalent to the
replacement of the inner product of Eq.(5.11) by the inner product of Eq.(5.20). After 
performing the Wick rotation we continue our analysis in a straightforward manner. For 
instance, the inner product of a vector ${\vec V}:=V^I{\hat e}_I\in\sigma_w$ with itself
is now:
\begin{equation}
{\vec V}\bullet{\vec V} = V^I V^J\eta_{IJ} = -(V^0)^2 + (V^1)^2 + (V^2)^2 + (V^3)^2,
\end{equation}
instead of being the one of Eq.(5.13).

  When performing the Wick rotation we introduce, simultaneously, the concept of {\it time}
into our model of spacetime. In general, the problems concerning the concepts of time and 
causality are the most profound ones in gravitational physics. In our model the concept of
time plays no role whatsoever at the Planck length scale, but time is brought into our model
from "outside", by means of the Wick rotation, as a macroscopic concept, without any 
connection with the microphysics of spacetime. This is certainly an unattractive feature of 
our model, but at this stage of research it is necessary, if we want to make any progress.

   At this point we define in $\sigma_w$ the vectors $\vec{b}_\mu$ such that
\begin{equation}
\vec{b}_\mu := \frac{1}{\lambda}\vec{v}_{4\mu}
\end{equation}
for all $\mu=0,1,2,3$. Using these vectors one may write the position vector $\vec{r}$ of an 
arbitrary point of $\sigma_w$ as:
\begin{equation}
\vec{r} = x^\mu\vec{b}_\mu,
\end{equation}
where, again, Einstein's sum rule has been used. The vectors $\vec{b}_\mu$ define new 
coordinates $x^\mu$ in $\sigma_w$. One finds that the coordinates $x^\mu$ have the property:
\begin{equation}
x^\mu(\tilde{v}_\nu) = \lambda\delta^\mu_\nu
\end{equation}
for every $\mu$,$\nu=0,1,2,3$. In other words, the coordinates of the vertex $\tilde{v}_0$ are 
$(\lambda,0,0,0)$, those of $\tilde{v}_1$ are $(0,\lambda,0,0)$, and so on. It is easy to 
see that when we go from the vertex $\tilde{w}$ to the vertex $\tilde{v}_\mu$ in $\sigma_w$,
the change $\Delta x^\mu$ in the coordinate $x^\mu$ in $\sigma_w$ is the same as is the 
change $\Delta\chi^\mu$ in the coordinate $\chi^\mu$ in $G_v$, when we go from the vertex 
$w$ to the vertex $v_\mu$ in $G_v$.

   The geometrical four-simplex $\sigma_w$ now plays in the vicinity of the vertex $w$ in 
$G_v$ a role similar to the one played by the tangent space $T_P$ in the vicinity of a point
$P$ in classical spacetime. The tangent space $T_P$ is a flat Minkowski spacetime, which is
spanned by the tangent vectors of the coordinate curves meeting at the point $P$, whereas
the four-simplex $\sigma_w$ is a flat Minkowski spacetime spanned by the vectors 
$\vec{b}_\mu$ which, in our model, play the role of the tangent vectors of the coordinate 
curves. Moreover, the tangent space $T_P$ approximates classical spacetime in the vicinity
of the point $P$ at length scales, where the classical general relativistic effects on 
spacetime curvature may be neglected, whereas the geometrical four-simplex $\sigma_w$ approximates quantum
spacetime in the vicinity of the vertex $w$ at length scales, where both the effects caused
by the Planck size quantum fluctuations as well as the effects caused by classical general
relativity on spacetime curvature may be ignored. The size of the four-simplex $\sigma_w$ is
determined by the number $\lambda\in(0,1)$. 

  Since the vectors $\vec{b}_\mu$ play the role of the tangent vectors of the coordinate 
curves in our model, we say that the vectors $\vec{b}_\mu$ constitute a {\it base determined
by the coordinates} $\chi^\mu$. In this base one may write the components of {\it tensors}
determined at the vertex $w$ with the coordinates $(\chi^0,\chi^1,\chi^2,\chi^3)$. For 
instance, the components of the {\it metric tensor} are:
\begin{equation}
g_{\mu\nu} := \vec{b}_\mu\bullet\vec{b}_\nu 
= \frac{1}{\lambda^2}\vec{v}_{4\mu}\bullet\vec{v}_{4\nu},
\end{equation}
where the last equality follows from Eq.(5.22). Eq.(5.19) implies that 
\begin{equation}
{\vec v}_{4\mu} = X^I_\mu{\hat e}_I,
\end{equation}
where $X^I_a$'s are the coordinates of the vertex ${\tilde v}_u$ for all $\mu=0,1,2,3$. 
Hence it follows that
\begin{equation}
g_{\mu\nu} = \frac{1}{\lambda^2} X^I_\mu X^J_\nu{\hat e}_I\bullet{\hat e}_J 
= \frac{1}{\lambda^2}X^I_\mu X^J_\nu\eta_{IJ}.
\end{equation}
In other words, if we know the coordinates $X^I_\mu$ of the vertices of the four-simplex
$\sigma_w$, we may calculate the metric tensor at the vertex $w$ of spacetime. The 
coordinates $X^I_\mu$, in turn, may be solved from Eqs.(5.17)-(5.19) in terms of the areas
$A_{abc}$ of the two-faces of $\sigma_w$.

  To see what all this means consider, as an example, a special case, where the areas of 
all two-faces are equal, i.e. $A_{abc}=A$ for all $a,b=0,1,2,3,4$. In that case all edge 
lengths are equal, and it is straightforward to show, using Eqs.(5.17)-(5.19), that
\begin{subequations}
\begin{eqnarray}
{\tilde v}_0 &=& (\frac{\sqrt{10}}{4}l, \frac{1}{2}l, 
\frac{\sqrt{3}}{6}l, \frac{\sqrt{6}}{12}l),\\
{\tilde v}_1 &=& (0,l,0,0),\\
{\tilde v}_2 &=& (0,\frac{1}{2}l,\frac{\sqrt{3}}{2}l,0),\\
{\tilde v}_3 &=& (0,\frac{1}{2}l,\frac{\sqrt{3}}{6}l,\frac{\sqrt{6}}{3}l),
\end{eqnarray}
\end{subequations}
where
\begin{equation}
l^2 := \frac{4\sqrt{3}}{3}A
\end{equation}
is the square of the common length $l$ of the edges of $\sigma_w$. Using Eq.(5.27) we 
therefore find that the components of the metric tensor $g_{\mu\nu}$ may be arranged into
the following matrix:
\begin{equation}
(g_{\mu\nu}) = \frac{\sqrt{3}}{3\lambda^2}A\left(\begin{array}{cccc}
-1&2&2&2\\
2&4&2&2\\
2&2&4&2\\
2&2&2&4\end{array}\right).
\end{equation}
It should be noted that the expression of Eq.(5.27) for the components of the metric tensor 
is independent of the choice of the coordinates $X^I$ inside the four-simplex $\sigma_w$. 
More precisely, the components of the metric tensor $g_{\mu\nu}$ are preserved invariant in
the Lorentz transformations and the rotations of the coordinates $X^I$. The only 
requirements posed for the coordinates $X^I$ are that, with respect to these coordinates, 
the three-simplex, or tetrahedron ${\tilde w}{\tilde v}_1{\tilde v}_2{\tilde v}_3$ must be
a spacelike three-space, and the vector ${\vec v}_{40}$ must be future pointing and 
timelike.

  Eq.(5.27) determines the metric structure of our spacetime model. $g_{\mu\nu}$ is a
function of the coordinates $\chi^\mu$ of the vertex $w$. In other words, $g_{\mu\nu}$
depends on the vertex $w$, where $g_{\mu\nu}$ has been written. One might think that 
$g_{\mu\nu}$ is also a function of the parameter $\lambda\in(0,1)$. It turns out, however,
that at macroscopic length scales $g_{\mu\nu}$ is a function of the coordinates $\chi^\mu$ 
only. The chain of reasoning producing this important conclusion runs as follows:

    The geometrical four-simplex $\sigma_w$ approximates spacetime in the vicinity of the
vertex $w$. In particular, the edges $\tilde{w}\tilde{v}_\mu$ approximate the corresponding
discrete geodesics in $G_v$. These discrete geodesics, in turn, approximate in $G_v$ the 
coordinate curves, where one of the indices $k_\mu$ changes, and three others are constants.
One finds from Eq.(5.6) that the number of edges in the coordinate curve which connects the
vertices $w$ and $v_\mu$ is directly proportional to $\lambda$. Since the discrete geodesics
$wv_\mu$ approximate the corresponding coordinate curves, one may expect that the numbers of
edges in the discrete geodesics $wv_\mu$ is also directly proportional to $\lambda$. When 
the number of edges is very large, the length of a discrete geodesic is directly
proportional to the number of its edges. Since the number of edges in the discrete geodesic
$wv_\mu$ is directly proportional to $\lambda$, it follows that its length is also 
proportional to $\lambda$. Futhermore, since the edge $\tilde{w}\tilde{v}_\mu$ of $\sigma_w$
approximates the discrete geodesic $wv_\mu$, we find that the length of the edge 
$\tilde{w}\tilde{v}_\mu$ is proportional to $\lambda$. Finally, since the numerical values 
of the coordinates $X^I_\mu$ of the vertex ${\tilde v}_\mu$ in Eq.(5.27) are directly
proportional to the length of the edge ${\tilde w}{\tilde v}_\mu$, we find that $g_{\mu\nu}$
is independent of $\lambda$. In other words, when $\lambda N^{1/4}>>1$, then $g_{\mu\nu}$ is
a function of the spacetime coordinates $\chi^\mu$ only.

  The metric tensor $g_{\mu\nu}$ is the fundamental object of classical general relativity,
and therefore we have shown, how classical spacetime emerges from quantum spacetime: The
areas of discrete triangles depend on the quantum states of the microscopic black holes at 
their vertices, and using Eqs.(5.17), (5.18) and (5.27) we find in which way $g_{\mu\nu}$ 
depends on these areas. In other words, we have reduced $g_{\mu\nu}$ to the quantum states 
of the microscopic quantum black holes, which constitute spacetime. However, our chain of 
reasoning, which produced the important result that $g_{\mu\nu}$ is a function of the 
spacetime coordinates $\chi^\mu$ only, hinged on the crucial assumption that {\it the length
of a discrete geodesic is directly proportional to the number of its edges}. Certainly this
assumption is valid only when the number of edges is enormous, i.e. at the macroscopic 
length scales. So it follows that $g_{\mu\nu}$ tells nothing about the microscopic physics
of spacetime. Rather, $g_{\mu\nu}$ is a sort of {\it statistical} or {\it thermodynamical}
quantity, which may be used for the description of spacetime properties at macroscopic
lenght scales only. 
 
   Keeping in mind that $g_{\mu\nu}$ is just a thermodynamical quantity of spacetime, one
may define, in terms of $g_{\mu\nu}$, other thermodynamical quantities familiar from 
classical general relativity. For instance, one may define the {\it Christoffel symbol}
\begin{equation}
\Gamma^\alpha_{\mu\nu} := \frac{1}{2}g^{\alpha\sigma}
(\frac{\partial g_{\sigma\mu}}{\partial\chi^\nu} 
+ \frac{\partial g_{\nu\sigma}}{\partial\chi^\mu} 
- \frac{\partial g_{\mu\nu}}{\partial\chi^\sigma}).
\end{equation}
When the number of vertices is enormous, the allowed values of $\chi^\mu$ constitute, in
practise, a continuous set of points, and the differences may be replaced, as an excellent
approximation, by differentials. Hence Eq.(5.31) really makes sense at macroscopic length
scales. For the same reason we may meaningfully talk about diffeomorphism symmetry at 
macroscopic length scales. By means of $\Gamma^\alpha_{\mu\nu}$ and $g_{\mu\nu}$ one may define, in the usual 
way, the Riemann and the Ricci tensors, curvature scalars, covariant derivatives, and so 
on. However, it should always be kept in mind that all these objects describe the {\it 
thermodynamics} of spacetime only.

\maketitle

\section{Thermodynamics of Spacetime}

   In the previous Section we found how the fundamental concepts of classical general
relativity arise as sort of thermodynamical quantities out of quantum spacetime. With
the concepts of classical general relativity in our service, although equipped with a 
new interpretation, we are now prepared to investigate the thermodynamics of spacetime.

   When investigating the thermodynamics of spacetime, the first task is to construct,
in the context of our model, the definitions for two fundamental concepts of 
thermodynamics. These fundamental concepts are {\it internal energy}, or {\it heat} of a 
system, and {\it temperature}. What do these concepts mean in quantum spacetime?

   When attempting to construct the definition of heat in quantum spacetime we can do 
nothing better than to seek for ideas and inspiration from the good old Newtonian theory of 
gravitation. This theory is based on Newton's universal law of gravitation, which states 
that point-like bodies attract each other with a gravitational force, which is directly
proportional to the masses of the bodies, and inversely proportional to the square of their
distance. This law implies that a point-like body with mass $M$ creates in its neighborhood
a gravitational field 
\begin{equation}
\vec{g}(\vec{r}) = -G\frac{M}{r^2}\hat{e}_r,
\end{equation}
where $\vec{r}$ is the position vector of the point in which the field is measured such that
the body under consideration lies at the origin of the system of coordinates. $r$ is the 
distance of that point from the body, and $\hat{e}_r$ is the unit vector parallel to 
$\vec{r}$. The gravitational field $\vec{g}(\vec{r})$ tells the {\it acceleration} an
observer at rest with respect to the body will measure, at the point $\vec{r}$, 
for all bodies in a free fall in the gravitational field created by the mass $M$. One finds
that if ${\cal S}$ is a closed, orientable two-surface, there is an interesting relationship
between the mass $M$, and the flux of the gravitational field $\vec{g}(\vec{r})$ through
that surface:
\begin{equation}
M = -\frac{1}{4\pi G}\oint_{\cal S} \vec{g}\bullet\hat{n}\,dA,
\end{equation}
where $\hat{n}$ is the outward pointing unit normal of that surface, and $dA$ is the area
element of the surface. It turns out that Eq.(6.2) holds not only for a single point-like
mass, but it holds for arbitrary mass distributions. As a generalization of Eq.(6.2) we may 
write:
\begin{equation}
M_{tot} = -\frac{1}{4\pi G}\oint_{\cal S} \vec{g}\bullet\hat{n}\,dA,
\end{equation}
where $M_{tot}$ is the total mass of the mass distribution inside the closed surface 
${\cal S}$. 
Since mass and energy are equivalent, one might expect that the right hand side of Eq.(6.3)
would provide, at least when the gravitational field is very weak, and the speeds of the
massive bodies very low, a some kind of notion of gravitational energy.

   The general relativistic generalization of Eq.(6.3) is obvious: In essense, we just 
replace the gravitational field $\vec{g}$, which tells the acceleration under presence
of the gravitating bodies, by the {\it proper acceleration}
\begin{equation}
a^\mu := \xi^\alpha\xi^\mu_{;\alpha}
\end{equation}
corresponding to the timelike Killing vector field $\xi^\alpha$ of spacetime. In Eq.(6.4)
the semicolon means covariant differentiation. We may define an integral 
\begin{equation}
E(V) := \frac{1}{4\pi}\oint_{\partial V} a^\mu n_\mu\,dA,
\end{equation}
where $V$ is a simply connected domain of a spacelike hypersurface of spacetime, 
$\partial V$ is its boundary, and $n_\mu$ is a spacelike unit normal vector of $\partial V$.
Another way of writing Eq.(6.5) is:
\begin{equation}
E(V) := \frac{1}{8\pi}\oint_{\partial V} \xi^{\mu;\nu}\,d\Sigma_{\mu\nu},
\end{equation}
where we have defined:
\begin{equation}
d\Sigma_{\mu\nu} := (n_\mu\xi_\nu - n_\nu\xi_\mu)\,dA.
\end{equation}
That the integrals on the right hand sides of Eqs.(6.5) and (6.6) are really the same
follows from the fact that $\xi^\mu$ obeys the Killing equation:
\begin{equation}
\xi_{\mu;\nu} + \xi_{\nu;\mu} = 0.
\end{equation}
The right hand side Eq.(6.6) is known as the {\it Komar integral} 
\cite{kaksikymmentayksi, kaksikymmentakaksi, kaksikymmentakolme}, and it gives a
satisfactory definition for the concept of energy in certain stationary spacetimes.

   As an example, one may consider the Schwarzschild spacetime, where:
\begin{equation}
ds^2 = -(1-\frac{2M}{r})\,dt^2 + \frac{dr^2}{1-\frac{2M}{r}} + r^2\,d\theta^2 
+ r^2\sin^2\theta\,d\theta^2.
\end{equation}
When $r>2M$, this spacetime admits a timelike Killing vector field $\xi^\mu$ such that the
only non-zero component of this vector field is:
\begin{equation}
\xi^t \equiv 1.
\end{equation}      
On the spacelike two-sphere, where $r=constant$, the only non-zero component of $n^\mu$ is:
\begin{equation}
n^r = -(1-\frac{2M}{r})^{1/2}.
\end{equation}
One finds that the Komar integral of Eq.(6.6) becomes:
\begin{equation}
E = (1-\frac{2M}{r})^{-1/2}M.
\end{equation}
This gives the energy of the gravitational field from the pont of view of an observer at
rest with respect to the Schwarzschild coordinate $r$. $(1-(2M)/r)^{-1/2}$ is the red-shift 
factor.

    The Komar integral provides a satisfactory definition of energy in {\it stationary} 
spacetimes. How to define the concept of energy in general, non-stationary spacetimes? In 
particular, how to define the concept of {\it heat?} 

    Unfortunately, it is impossible to find a satisfactory definition of energy, or even 
energy density, in non-stationary spacetimes. (For a detailed discussion of this problem,
see Ref.\cite{kuusi}.). However, it might be possible to attribute meaningfully the concept of heat 
to some specific spacelike two-surfaces of spacetime. After all, we found in Section 4 that 
spacelike two-surfaces possess entropy. If they possess entropy, then why should they not 
possess, in some sense, heat as well?

   Valuable insights into this problem are provided by the investigations we made above 
about the properties of Newtonian gravity and Komar integrals. Those investigations suggest
that when we attempt to construct a physically sensible definition of heat of spacelike
two-surfaces, the flux
\begin{equation}
\Phi := \int_{\cal S} a^\mu n_\mu\,dA
\end{equation}
of the proper acceleration vector field
\begin{equation}
a^\mu := u^\alpha u^\mu_{;\alpha}
\end{equation}
through the two-surface under consideration might play an important role \cite{lisakaksi}.
 After all, both in
Eqs.(6.3) and (6.5) we calculated the flux of an acceleration vector field $a^\mu$ through a certain 
closed, spacelike two-surface. However, there is an important difference between the 
definitions (6.4) and (6.14) of the vector field $a^\mu$: In Eq.(6.4) the vector field
$a^\mu$ was defined by means of an appropriately chosen Killing vector field $\xi^\mu$, 
whereas in Eq.(6.14) $a^\mu$ is defined by means of a future pointing unit tangent vector
field $u^\mu$ of the congruence of the timelike world lines of the points of an arbitrary
spacelike two-surface of spacetime.

    In our investigations concerning the thermodynamics of spacetime we shall focus our
attention at those spacelike two-surfaces, where the proper acceleration vector field $a^\mu$
of the congruence of the world lines of the points of the two-surface has the following 
properties:
\begin{subequations}
\begin{eqnarray}
\sqrt{a^\mu a_\mu} = constant := a,\\
a^\mu n_\mu = a.
\end{eqnarray}
\end{subequations}
at every point of the two-surface (In this Section we consider spacetime at macroscpic 
lenght scales, and therefore we may ignore its discrete structure.). In other words, 
we shall assume that all points of the spacelike two-surface under consideration have 
all the time the same constant proper acceleration $a$, and every point of the 
two-surface is accelerated to the direction orthogonal to the two-surface. For the sake 
of brevity and simplicity we shall call such spacelike two-surfaces as {\it acceleration
surfaces}. It is easy to see that the flux of the proper acceleration vector field through 
an acceleration surface is
\begin{equation}
\Phi_{as} = aA,
\end{equation}
where $A$ is the area of the acceleration surface.

   The motivation for our definition of the concept of acceleration surface is provided 
by the fact that acceleration surfaces are very similar to the event horizons of black
holes: The surface gravity $\kappa$ is constant everywhere and all the time on a black hole
event horizon, whereas on an acceleration surface the proper acceleration $a$ is a constant.
For black hole event horizons one may meaningfully associate the concepts of heat, entropy
and temperature, and there are good hopes that the same might be done for acceleration 
surfaces as well.

    Motivated by the similarities between black hole event horizons and acceleration 
surfaces, as well as by the properties of Komar integrals, we now define the {\it change
of heat} of an acceleration surface in terms of the differential $d\Phi_{as}$ of the flux 
$\Phi_{as}$ of the proper acceleration vector field through the acceleration surface as:
\begin{equation}
\delta Q_{as} := \frac{1}{4\pi}\,d\Phi_{as}
\end{equation}
or, in SI units:
\begin{equation}
\delta Q_{as} := \frac{c^2}{4\pi G}\,d\Phi_{as}.
\end{equation}
As such as it is, however, this is just an empty definition, and several questions arise:
What is the physical interpretation of $\delta Q_{as}$? What are its physical effects? How
to measure $\delta Q_{as}$?

  To find an answer to these questions, consider thermal radiation flowing through an 
acceleration surface. We parametrize the world lines of the points of the surface by means 
of the proper time $\tau$ measured along those world lines. We shall assume that the 
acceleration surface has a property:
\begin{equation}
\frac{\delta Q_{as}}{d\tau}\vert_{\tau=0} = 0,
\end{equation}
i.e. when $\tau=0$, the rate of change in the heat content of the surface is zero. An 
example of this kind of an acceleration surface is a surface, where $a$ is constant in time,
and $\frac{dA}{d\tau}\vert_{\tau=0}=0$. When radiation flows through the acceleration 
surface, heat and entropy are carried through the surface, and presumably the radiation 
interacts with the surface such that its geometry is changed. For instance, the area of 
the surface may change. However, if the geometry of the acceleration surface changes, so 
does its heat content, and the heat delivered or absorbed by the surface contributes to
the flow $\frac{\delta Q_{rad}}{d\tau}$ of the heat $Q_{rad}$ carried by the radiation 
through the surface. As a result $\frac{\delta Q_{rad}}{d\tau}$ changes in the proper time
$\tau$, and we must have
\begin{equation}
\frac{\delta^2 Q_{rad}}{d\tau^2}\vert_{\tau=0} \ne 0,
\end{equation}
where $\frac{\delta^2 Q_{rad}}{d\tau^2}$ denotes the rate of change of 
$\frac{\delta Q_{rad}}{d\tau}$ with respect to the proper time $\tau$. The quantities 
$\frac{\delta Q_{rad}}{d\tau}$ and $\frac{\delta^2 Q_{rad}}{d\tau^2}$ have been measured
from the point of view of an observer at rest with respect to the acceleration surface.

  Conservation of energy now implies that if Eq.(6.19) holds, then the rate of increase 
in the flow of heat carried by radiation through the acceleration surface is the same as 
is the decrease in the rate of change in the heat content of the surface. In other words, 
the heat of the acceleration surface is exactly converted to the heat of the radiation,
and vice versa. A mathematical expression for this statement is:
\begin{equation}
\frac{\delta^2 Q_{rad}}{d\tau^2}\vert_{\tau=0} 
= -\frac{\delta^2 Q_{as}}{d\tau^2}\vert_{\tau=0}.
\end{equation}
In our model, this equation is the fundamental equation of the thermodynamics of spacetime.
According to the definition (6.17), Eq.(6.21) implies:
\begin{equation}
\frac{\delta^2 Q_{rad}}{d\tau^2}\vert_{\tau=0} 
= -\frac{1}{4\pi}\frac{d^2\Phi_{as}}{d\tau^2}\vert_{\tau=0}
\end{equation}
or, in SI units:
\begin{equation}
\frac{\delta^2 Q_{rad}}{d\tau^2}\vert_{\tau=0}
= -\frac{c^2}{4\pi G}\frac{d^2\Phi_{as}}{d\tau^2}\vert_{\tau=0}.
\end{equation}

  Consider now the possible implications of Eq.(6.21). As the first example, consider a 
very small plane, which is in a uniformly accelerating motion with a constant proper 
acceleration $a$ to the direction of its spacelike unit normal vector. Obviously, such a 
plane is an acceleration surface, and we may assume that our surface statisfies Eq.(6.19).
Assuming that spacetime is filled with radiation in thermal equilibrium, we find that 
Eq.(6.21) implies:
\begin{equation}
\frac{\delta^2 Q_{rad}}{d\tau^2}\vert_{\tau=0} 
= -\frac{a}{4\pi}\frac{d^2 A}{d\tau^2}\vert_{\tau=0},
\end{equation}
where $A$ is the area of the plane such that $\frac{dA}{d\tau}\vert_{\tau=0}=0$. It follows
from Eq.(1.2) which, moreover, follows from the first law of thermodynamics, that
\begin{equation}
\delta Q_{rad} = T_{rad}\,dS_{rad},
\end{equation}
where $dS_{rad}$ is the amount of entropy carried by radiation out of the plane, and 
$T_{rad}$ is its temperature. Assuming that $T_{rad}$ is constant during the process, we
may write Eq.(6.24) as:
\begin{equation}
T_{rad}\frac{d^2 S_{rad}}{d\tau^2}\vert_{\tau=0} 
= -\frac{a}{4\pi}\frac{d^2 A}{d\tau^2}\vert_{\tau=0}.
\end{equation}
At this point we recall Eq.(4.9), which implies that every spacelike two-surface of 
spacetime has an entropy, which is proportional to the area of that two-surface. Using 
that equation we find that
\begin{equation}
\frac{d^2 A}{d\tau^2}\vert_{\tau=0} 
= \frac{\alpha}{\ln 2}\frac{d^2 S_{plane}}{d\tau^2}\vert_{\tau=0},
\end{equation}
where $S_{plane}$ denotes the entropy content of the plane. Hence, Eq.(6.26) takes the form:
\begin{equation}
T_{rad}\frac{d^2 S_{rad}}{d\tau^2}\vert_{\tau=0} 
= -\frac{a}{4\pi}\frac{\alpha}{\ln 2}\frac{d^2 S_{plane}}{d\tau^2}\vert_{\tau=0}.
\end{equation}
This equation allows us to associate the concept of {\it temperature} with our accelerating 
plane: When the temperatures of the radiation and the plane are equal, the entropy loss of 
the plane is equal to the entropy gain of the radiation. In other words, the entropy of the
plane is exactly converted to the entropy of the radiation. In this case the plane is in a 
thermal equilibrium with the radiation, and we have:
\begin{equation}
\frac{d^2 S_{plane}}{d\tau^2}\vert_{\tau=0} = -\frac{d^2 S_{rad}}{d\tau^2}\vert_{\tau=0},
\end{equation}
and Eq.(6.28) implies:
\begin{equation}
T_{rad} = \frac{\alpha}{\ln 2}\frac{a}{4\pi}.
\end{equation}
As one may observe, in thermal equilibrium the temperature of the radiation is proportional
to the proper acceleration $a$ of the plane.

   Now, how should we interpret this result? A natural interpretation is that an 
accelerating observer observes thermal radiation with a characteristic temperature, which 
is directly proportional to his proper acceleration. Actually, this is a well known result 
of relativistic quantum field theories, and it is known as the {\it Unruh effect} 
\cite{kaksikymmentanelja}. According to 
this effect an accelerating observer observes thermal particles even when, from the point
of view of all inertial observers, there are no particles at all. The characteristic 
temperature of the thermal particles is the so called {\it Unruh temperature}
\begin{equation}
T_U := \frac{a}{2\pi}
\end{equation}
or, in SI units:
\begin{equation}
T_U := \frac{\hbar a}{2\pi k_B c}.
\end{equation}
Comparing Eqs.(6.30) and (6.31) we find that the temperature $T_{rad}$ predicted by our 
model for the thermal radiation equals to the Unruh temperature $T_U$, provided that           
\begin{equation}
\alpha = 2\ln 2.
\end{equation}
It is most gratifying that our quantum mechanical model of spacetime predicts the Unruh
effect. According to our model the Unruh effect is a direct outcome of the statistics of
spacetime.

    Eq.(6.33), together with Eq.(4.9), implies that, in the low-temperature limit, the
entropy of an arbitrary spacelike two-surface of spacetime is, in natural units,
\begin{equation}
S = \frac{1}{2}A
\end{equation}
or, in SI units,
\begin{equation}
S = \frac{1}{2}\frac{k_B c^3}{\hbar G}A.
\end{equation}
In other words, our model predicts that, in the low-temperature limit, every spacelike
two-surface of spacetime has entropy which, in natural units, is one-half of its area. 
This result is closely related to the famous {\it Bekenstein-Hawking entropy law},
which states that black hole has entropy which, in natural units, is one-quarter of its
event horizon area \cite{kolme, nelja}. In other words, our model predicts for the entropy of a spacelike
two-surface a numerical value, which is exactly {\it twice} the numerical value of the
entropy of a black hole event horizon with the same area. At this point it should be 
noted that we would also have been able to obtain the Unruh temperature of Eq.(6.31) 
for an accelerating plane straightforwardly from the first law of thermodynamics, and an 
assumption that the plane has an entropy which is one-half of its area: It follows from 
Eqs.(6.16), (6.17) and (6.34) that if the proper acceleration $a$ on an acceleration 
surface is kept as a constant, then an infinitesimal change $\delta Q$ in its heat may
be expressed in terms of an infinitesimal change $dS$ in its entropy as:
\begin{equation}
\delta Q = \frac{a}{2\pi}\,dS,
\end{equation}
which readily implies that the temperature of the surface is the Unruh temperature $T_U$ of
Eq.(6.31). 

   One of the consequences of the Bekenstein-Hawking entropy law is the
{\it Hawking effect}: Black hole emits thermal particles with the characteristic
temperature
\begin{equation}
T_H := \frac{\kappa}{2\pi},
\end{equation}
which is known as the {\it Hawking temperature} \cite{nelja}. In Eq.(6.37) $\kappa$ is the surface
gravity at the horizon of the hole. For a Schwarzschild black hole with mass $M$ we have
\begin{equation}
T_H = \frac{1}{8\pi M}
\end{equation}
or, in SI units:
\begin{equation}
T_H = \frac{\hbar c^3}{8\pi G k_B}\frac{1}{M}.
\end{equation}
The Hawking temperature gives the temperature of the black hole radiation from the point
of view of a faraway observer at rest with respect to the hole, provided that the
backreaction and the backscattering effects are neglected. If the observer lies at a finite
(although very small) distance from the event horizon of the hole, Eq.(6.38) must be
corrected by the red shift factor, and we get:
\begin{equation}
T_H = \frac{1}{8\pi}(1-\frac{2M}{r})^{-1/2}\frac{1}{M}.
\end{equation}

  Consider now how Eq.(6.40), and hence the Hawking effect for the Schwarzschild black hole,
may be obtained from our model. Our derivation of Eq.(6.40) from our model of spacetime will
also bring some light to the curious fact that the entropy of a spacelike two-surface is,
according to our model, exactly twice the entropy of a black hole event horizon with the
same area. 

  The only non-zero component of the future pointing unit tangent vector $u^\mu$ of an
observer at rest with respect to the Schwarzschild coordinates $r$ and $t$ is
\begin{equation}
u^t = (1-\frac{2M}{r})^{-1/2},
\end{equation}
and the only non-zero component of the corresponding four-acceleration $a^\mu$ is:
\begin{equation}
a^r = u^t u^r_{;t} = -\frac{M}{r^2}.
\end{equation}
It is easy to see that the spacelike two-sphere, where $r=constant(>2M)$ is an acceleration
surface
Using Eqs.(6.11) and (6.14) we find that the flux of the vector field $a^\mu$ through that 
two-sphere is:
\begin{equation}
\Phi_{as} = 4\pi M(1-\frac{2M}{r})^{-1/2},
\end{equation}
where we have used the fact that the area of the two-sphere, where $r=constant$ is:
\begin{equation}
A = 4\pi r^2.
\end{equation}
Using Eq.(6.34) we find that in the low temperature limit the entropy of that two-sphere is:
\begin{equation}
S = 2\pi r^2.
\end{equation}

  When we obtained the Unruh effect from our model, we varied the flux $\Phi_{as}$ in Eq.(6.36)
in such a way that we kept the proper acceleration $a$ as a {\it constant}, and varied the
area $A$ only. The proper acceleration $a$ then took the role of temperature, and the area
$A$ that of entropy. We shall now use the same idea, when we obtain the Hawking effect
from our model: When varying the flux $\Phi_{as}$ of Eq.(6.43) we keep the quantity
\begin{equation}
a := a^\mu n_\mu = (1-\frac{2M}{r})^{-1/2}\frac{M}{r^2}
\end{equation}
as a constant. In other words, we consider $a$ as a function of both $M$ and $r$, and we 
require that the total differential of $a$ vanishes:
\begin{equation}
da = \frac{\partial a}{\partial M}\,dM + \frac{\partial a}{\partial r}\,dr = 0,
\end{equation}
which implies:
\begin{equation}
dM = \frac{2Mr-3M^2}{r^2-Mr}\,dr,
\end{equation}
i.e. an infinitesimal change $dM$ in the Schwarzschild mass $M$ of the hole must be 
accompained with a certain change $dr$ in the radius $r$ of the two-sphere. Using Eqs.(6.43)
and (6.48) one finds that when $a$ is kept constant, then the change $\delta Q$ in the heat 
content of the two-sphere may be written in terms of $dr$ as:
\begin{equation}
\delta Q = \frac{1}{4\pi}\,d\Phi = \frac{2M}{r}(1-\frac{2M}{r})^{-1/2}\,dr,
\end{equation}
and because Eq.(6.45) implies that the corresponding maximum  change in the entropy of the
two-sphere is
\begin{equation}
dS = 4\pi r\,dr,
\end{equation}
we find the relationship between $\delta Q$ and $dS$:
\begin{equation}
\delta Q = \frac{M}{2\pi r^2}(1-\frac{2M}{r})^{-1/2}\,dS.
\end{equation}
Therefore, according to the first law of thermodyamics, the two-sphere has a temperature
\begin{equation}
T = \frac{M}{2\pi r^2}(1-\frac{2M}{r})^{-1/2}.
\end{equation}
Hence we find that an observer with constant $r$ just outside the horizon, where $r=2M$, 
will observe that the black hole has a temperature
\begin{equation}
T_H = \frac{1}{8\pi}(1-\frac{2M}{r})^{-1/2}\frac{1}{M},
\end{equation}
which is Eq.(6.40). So we have shown that the Hawking effect is just one of the consequences
of our model in the low temperature limit.

  Let us now investigate our derivation of the Hawking effect in more details. The crucial
points in our derivation were our decisions to consider a two-sphere just outside the
horizon, instead of the horizon itself, and to keep $a^\mu n_\mu$ as a constant while 
calculating the infinitesimal change in the heat content of the two-sphere. In these
points our derivation differed from the usual derivations of the Hawking effect: In the
usual derivations one uses as the starting point the so called {\it mass formula of
black holes}, which is sometimes also known as the Smarr formula \cite{kaksikymmentakaksi}.
For Schwarzschild black
holes this formula implies that the Schwarzschild mass $M$ of the Schwarzschild black hole may be written in terms of the surface gravity $\kappa$ at
the horizon, and the horizon area $A_h$ as:
\begin{equation}
M = \frac{1}{4\pi}\kappa A_h.
\end{equation}
The surface gravity $\kappa$, which may be written in terms of $M$ as:
\begin{equation}
\kappa = \frac{1}{4M},
\end{equation}
may be viewed as an analogue of the quantity $a^\mu n_\mu$ in the sense that $\kappa$ gives
the proper acceleration of an object in a free fall at the horizon from the point of view
of a faraway observer at rest with respect to the hole. From the point of view of an
observer at rest just outside the event horizon of the hole the absolute value of the
proper acceleration of objects in a free fall just outside the hole is:
\begin{equation}
a = (1-\frac{2M}{r})^{-1/2}\kappa
\end{equation}
which, according to Eqs.(6.46) and (6.55), is exactly $a^\mu n_\mu$.

   Consider now what happens when we vary the right hand side of Eq.(6.54) in such a way
that during the variation we are all the time {\it at the horizon}. In that case $\kappa$ is
{\it not constant}, but it also varies such that we have:
\begin{equation}
dM = \frac{1}{4\pi}A_h\,d\kappa + \frac{1}{4\pi}\kappa\,dA_h,
\end{equation}
where \cite{kaksikymmentakaksi, kaksikymmentaviisi}
\begin{equation}
A_h\,d\kappa = -\frac{1}{2}\,dA_h.
\end{equation}
So we get:
\begin{equation}
dM = \frac{1}{8\pi}\kappa\,dA_h,
\end{equation}
which is the first law of black hole mechanics. Identifying, as usual, $\frac{1}{4}\,dA_h$ as
the change in the entropy of the hole, and $dM$ as the change in its heat content, we get
Eq.(6.38). So we find that the reason why the black hole entropy may be thought to be 
one-quarter, instead of one-half, of the horizon area, is that when the Schwarzschild mass
$M$ of the hole decreases as a result of the black hole radiance, the event horizon of the hole
shrinks such that the surface gravity $\kappa$ changes in the manner described in Eq.(6.58).
If the right hand side of Eq.(6.54) were varied in such a way that $\kappa$ is kept 
as a constant, then the entropy change corresponding to the area change $dA_h$ at the
temperature $T_H$ of Eq.(6.38) would be one-half, instead of one-quarter of $dA_h$.

\maketitle 

\section{Classical Limit: Einstein's Field Equation}

 We saw in the previous Section, much to our satisfaction, that our quantum mechanical model
of spacetime reproduces, in the low temperature limit, both the Unruh and the Hawking
effects. In other words, our model reproduces the well known semiclassical effects of 
gravity. 

   A really interesting question, and indeed the crucial test for our model, is whether the 
model implies, in the classical limit, Einstein's field equation. If it does, then we may 
say that Einstein's classical general relativity with all of its predictions is just one of
the consequences of our model.

   In this Section we show that Einstein's field equation indeed follows from our model in 
the classical limit. Our derivation will be based on the thermodynamical properties of 
spacetime, which were considered in the previous Section. It turns out that Einstein's field
equation is a simple and straightforward consequence of Eq.(6.21), the fundamental equation
of the thermodynamics of spacetime in our model. As in the previous Section, we consider
spacetime at length scales very much larger than the Planck lenght scale. At these lenght
scales we may consider spacetime, in effect, as a smooth (pseudo-) Riemannian manifold.
A detailed derivation of Einstein's field equation by means of thermodynamical 
considerations was performed in Ref.\cite{kaksikymmentakuusi}, and in this Section we shall
follow closely the ideas presented in that paper. 

     In Ref.\cite{kaksikymmentakuusi} an arbitrary point $P$ was picked up from spacetime, and
Einstein's field equation was obtained in that point. Since we are looking spacetime
at macroscopic length scales, the point $P$ has a neighborhood $U_P$, which may be 
equipped with a time orthogonal, or Gaussian normal system of coordinates, where the 
spacetime metric takes the form:  
\begin{equation}
ds^2 = -dt^2 + q_{ab}\,dx^a\,dx^b,
\end{equation}
and which becomes, at the point $P$, an orthonormal geodesic system of coordinates. In 
Eq.(7.1) $q_{ab}$ $(a,b=1,2,3)$ is the metric induced on the spacelike hypersurfaces, where
the time coordinate $t=constant$. Since our system of coordinates becomes an orthonormal 
geodesic system of coordinates at the point $P$, we have $q_{ab}=\delta_{ab}$ at that point.

  An arbitrary congruence of timelike curves at $U_P$ may be parametrized by means of the
time coordinate $t$ of Eq.(7.1). In other words, an arbitrary point $x^\mu$ of an arbitrary 
element of such a congruence may be expressed as a function of $t$. In 
Ref.\cite{kaksikymmentakuusi} 
it was shown that the tangent vector field  
\begin{equation}
w^\mu := \frac{dx^\mu}{dt}.
\end{equation}
of our congruence corresponding to this parametrization obeys an equation:
\begin{equation}
\frac{d\theta}{dt} = C^\mu_{;\mu} - w^\mu_{;\nu} w^\nu_{;\mu} + R_{\mu\nu}w^\mu w^\nu,
\end{equation}
where $R_{\mu\nu}$ is the Ricci tensor, and we have defined the scalar $\theta$ and the 
vector field $C^\mu$ such that:
\begin{subequations}
\begin{eqnarray}
\theta &:=& w^\mu_{;\mu},\\
C^\mu &:=& w^\alpha w^\mu_{;\alpha}.
\end{eqnarray}
\end{subequations}
Eq.(7.3) is the key equation in our derivation of Einstein's field equation from our 
model by means of thermodynamical considerations. The proof of Eq.(7.3) is easy, and it 
is very similar to the proof of the Raychaudhuri equation: Because
\begin{equation}
w^\alpha w_{\mu;\nu;\alpha} = w^\alpha w_{\mu;\alpha;\nu} - w^\alpha(w_{\mu;\alpha;\nu}
- w_{\mu;\nu;\alpha}),
\end{equation}
we get, using the product rule of covariant differentiation, and the basic properties of
the Riemann tensor $R^\alpha_{\,\,\beta\mu\nu}$:
\begin{equation}
w^\alpha w_{\mu;\nu;\alpha} = (w^\alpha w_{\mu;\alpha})_{;\nu} 
- w^\alpha_{;\nu}w_{\mu;\alpha}
+ R_{\beta\mu\nu\alpha}w^\alpha w^\beta.
\end{equation}
If we contract the indices $\mu$ and $\nu$, use the symmetry properties of the Riemann 
tensor, and rename the indices, we readily arrive at Eq.(7.3).

    As an example of a congruence of timelike curves parametrized by $t$, consider the 
special case, where spacetime is 
flat, and all points of all elements of the congruence are accelerated with a constant
proper acceleration $a$ to the direction of the negative $z$-axis. The points of the
elements of that kind of congruence obey an equation:
\begin{equation}
(z-z_0)^2 - (t-t_0)^2 = \frac{1}{a^2},
\end{equation}
where $t_0$ and $z_0$ are constants. Hence we observe that the only coordinates having 
explicit $t$-dependence are:
\begin{subequations}
\begin{eqnarray}
x^0 &=& t,\\
x^3 &=& z = -\sqrt{\frac{1}{a^2} + (t-t_0)^2} + z_0,
\end{eqnarray}
\end{subequations}
and therefore the only non-zero components of the vector field $w^\mu$ are:
\begin{subequations}
\begin{eqnarray}
w^0 &\equiv& 1,\\
w^3 &=& \frac{t_0-t}{\sqrt{\frac{1}{a^2} + (t-t_0)^2}},
\end{eqnarray}
\end{subequations}
and the only non-zero component of the vector field $C^\mu$ is:
\begin{equation}
C^3 = -\frac{2(t-t_0)^2 + \frac{1}{a^2}}{\lbrack\frac{1}{a^2} + (t-t_0)^2\rbrack^{3/2}}.
\end{equation}
One easily finds that the scalar $\theta$ vanishes identically, as well as do the first two 
terms on the right hand side of Eq.(7.3). Moreover, since $R_{\mu\nu}\equiv 0$ in flat 
spacetime, we find that Eq.(7.3) is indeed satisfied by the tangent vector field $w^\mu$
of our congruence.

  As in Ref.\cite{kaksikymmentakuusi}, we base our derivation of Einstein's field equation by means 
of thermodynamical arguments on a consideration of the properties of a very small spacelike 
two-plane in a uniformly accelerating motion. We parametrize the world lines of the points
of the plane by means of the proper time $\tau$ measured along those world lines. At the
point $P$ the plane is assumed to be parallel to the $xy$-plane, and in a uniformly 
accelerating motion with a constant proper acceleration $a$ to the direction of the negative
$z$-axis with respect to the orthonormal geodetic system of coordinates associated with the
point $P$. When the plane lies at the point $P$, the proper time $\tau=0$. When $\tau>0$, 
the plane is represented by the set of points, where $\tau=constant$. If spacetime is 
curved, the area of the spacelike two-surface, where $\tau=constant>0$, may be different 
from the area of the two-plane at the point $P$. In other words, the area of the plane 
may change.

   The possible change in the area of our accelerating two-plane, when it moves away from
the point $P$, may be calculated by means of Eq.(7.3). To begin with, we note that the 
change in the area $A$ of the two-plane during an infinitesimal time interval $dt$ is
\begin{equation}
dA = A(w^1_{;1} + w^2_{;2})\,dt
\end{equation}
provided that we are very close to the point $P$. $w^\mu$ is now the tangent vector field 
of the world lines of the points of the plane. It follows from Eq.(7.9)
that $w^1_{;1}=w^2_{;2}=w^3_{;3}=0$, i.e. $\frac{dA}{dt}=0$ at the point $P$. However, when 
the plane is moved away from the point $P$, then $\frac{dA}{dt}$ will become non-zero.

    Eq.(7.9), together with Eq.(7.1), implies that $w^0_{;0}\equiv 0$, and therefore we
may write, in general:
\begin{equation}
\theta = w^1_{;1} + w^2_{;2} + w^3_{;3}.
\end{equation}
Because $\theta$, together with the first two terms on the right hand side of Eq.(7.3) 
vanishes at the point $P$, Eq.(7.3), together with the chain rule, implies that after a very
short proper time interval $\tau$ we have
\begin{equation}
\theta = R_{\mu\nu} w^\mu w^\nu\frac{dt}{d\tau}\,\tau,
\end{equation}
where we have neglected the terms non-linear in $\tau$. Because between the tangent vector
field $w^\mu$, and the unit tangent vector field 
\begin{equation}
u^\mu := \frac{dx^\mu}{d\tau},
\end{equation}
of the congruence of the world lines of the points of the plane there is the relationship
\begin{equation}
w^\mu = u^\mu\frac{d\tau}{dt},
\end{equation}
we find that Eq.(7.13) may be written as:
\begin{equation}
\theta = R_{\mu\nu}u^\mu u^\nu\frac{d\tau}{dt}\tau.
\end{equation}

    The area change of our plane now depends on the properties of $w^3_{;3}$. If spacetime 
is isotropic, i.e. spacetime expands and contracts in exactly the same ways in all spatial 
directions at the point $P$, then $w^1_{;1}$, $w^2_{;2}$ and $w^3_{;3}$ are equals, and
it follows from the chain rule and Eqs.(7.11)-(7.16) that
\begin{equation}
\frac{d^2A}{d\tau^2} = \frac{2}{3}AR_{\mu\nu}u^\mu u^\nu
\end{equation}
at the point $P$. Another possibility worth of consideration is that spacetime expands and 
contracts along the $xy$ plane only. In that case $w^3_{;3}$ vanishes identically in the 
neigborhood of the point $P$, and we may write:
\begin{equation}
\frac{d^2A}{d\tau^2} = AR_{\mu\nu}u^\mu u^\nu
\end{equation} 
at the point $P$. Both of the Eqs.(7.17) and (7.18) will be used in our forthcoming 
investigations.

    So far we have considered the properties of the geometry of spacetime only. We shall
now turn our attention to the properties of matter. When a spacelike two-plane moves in 
spacetime, matter, and thereby energy and entropy, will flow through the plane. If the 
energy momentum stress tensor of the matter fields is $T_{\mu\nu}$, the energy flow through the
plane (energy flown through the plane in unit time) is:
\begin{equation}
\frac{dE_b}{d\tau} = AT_{\mu\nu}u^\mu n^\nu,
\end{equation}
where $n^\mu$ is the spacelike unit normal vector of the plane. We have denoted the energy
flow through the plane to the direction of the vector $n^\mu$ by $\frac{dE_b}{d\tau}$, 
because that energy flow is due to the movement of the plane with respect to the matter 
fields. In other words, Eq.(7.19) gives the {\it boost energy flow} through the plane.

    When the plane is accelerated with a constant proper acceleration $a$ to the direction
of the vector $-n^\mu$, the vectors $u^\mu$ and $n^\mu$ will change. After a very short 
proper time interval $\tau$ the vectors $u^\mu$ and $n^\mu$ have changed to the vectors 
$u'^\mu$ and $n'^\mu$ such that
\begin{subequations}
\begin{eqnarray}
u'^\mu &=& u^\mu + a\tau\, n^\mu,\\
n'^\mu &=& a\tau\, u^\mu + n^\mu.
\end{eqnarray}
\end{subequations}
We have assumed that $a\tau<<1$, and therefore we have neglected the terms non-linear in
$\tau$. Using Eq.(7.19) we find that under the assumption that
\begin{equation}
\frac{dA}{d\tau} = 0
\end{equation}
at the point $P$, the rate of change in the boost energy flow is, at the point $P$:
\begin{equation}
\frac{d^2E_b}{d\tau^2} = \frac{d^2E_{b,t}}{d\tau^2} + \frac{d^2E_{b,a}}{d\tau^2},
\end{equation}
where we have denoted:
\begin{subequations}
\begin{eqnarray}
\frac{d^2E_{b,t}}{d\tau^2} &:=& A\dot{T}_{\mu\nu}u^\mu n^\nu = AT_{\mu\nu,\alpha}u^\alpha
u^\mu n^\nu,\\
\frac{d^2E_{b,a}}{d\tau^2} &:=& aT_{\mu\nu}(u^\mu u^\nu + n^\mu n^\nu),
\end{eqnarray}
\end{subequations}
where the dot means proper time derivative. The first term on the righ hand side of 
Eq.(7.22) is due to the simple fact that the tensor $T_{\mu\nu}$ is different in different
points of spacetime, whereas the second term is due to the change of the velocity of the 
plane with respect to the matter fields. In other words, it is caused by the 
{\it acceleration} of the plane. In what follows, we shall focus our attention to that term.
If the proper acceleration $a$ of the plane is very large, the second term will dominate 
over the first term, and we may neglect the first term.

   We stated in the beginning of this Section that Einstein's field equation is a 
straightforward consequence of Eq.(6.21), the fundamental equation of the thermodynamics 
of spacetime in our model. That equation, however, was written for the {\it radiation}
which flows through an acceleration surface of spacetime. Because of that we shall first 
derive Einstein's field equation in the special case, where matter consists of massless,
non-interacting radiation (electromagnetic radiation, for instance) in thermal equilibrium.
The energy density of such radiation is, from the point of view of an observer at rest 
with respect to our plane: 
\begin{equation}
\rho = T_{\mu\nu}u^\mu u^\nu,
\end{equation}
and its pressure is
\begin{equation}
p = \frac{1}{3}\rho = T_{\mu\nu} n^\mu n^\nu.
\end{equation}
It is a specific property of this kind of matter that the tensor $T_{\mu\nu}$ is {\it
traceless}, i.e.
\begin{equation}
T^\alpha_\alpha = 0.
\end{equation}
Using Eq.(7.23b) we therefore find that
\begin{equation}
\frac{d^2E_{b,a}}{d\tau^2} = \frac{4}{3}aA\rho = \frac{4}{3}aAT_{\mu\nu}u^\mu u^\nu.
\end{equation}

     Now, it is easy to see that the right hand side of Eq.(7.27) gives the rate of change in 
the flow of {\it heat} through our accelerating plane, provided that all change in the flow
of boost energy is, in effect, caused by the mere acceleration of the plane. This
important conclusion follows from the well known fact that the entropy density (entropy
per unit volume) of the electromagnetic (or any massless, non-interacting) radiation in
thermal equilibrium is \cite{kaksikymmentayhdeksan}
\begin{equation}
s_{rad} = \frac{1}{T_{rad}}\frac{4}{3}\rho,
\end{equation}
where $T_{rad}$ is the absolute temperature of the radiation. One easily finds, by using the
first law of thermodynamics, that the rate of change in the flow of heat through our 
accelerating plane is
\begin{equation}
\frac{\delta^2 Q_{rad}}{d\tau^2} = \frac{4}{3}aA\rho,
\end{equation}
which is exactly Eq.(7.27).

    It is now very easy to obtain Einstein's field equation. We just use Eq.(6.24) which, in
turn, follows from Eq.(6.21). We shall assume that at the point $P$ our plane is at rest
with respect to the radiation, i.e.
\begin{equation}
T_{\mu\nu}u^\mu n^\nu = 0.
\end{equation}
In that case there is no net flow of radiation through the plane, and because radiation is 
assumed to be in thermal equilibrium, and therefore homogeneous and isotropic, spacetime 
expands and contracts in the same ways in all spatial directions. Hence we may use Eq.(7.17).
Eqs.(7.17), (7.27) and (6.24) imply:
\begin{equation}
\frac{4}{3}aAT_{\mu\nu}u^\mu u^\nu = -\frac{a}{6\pi}AR_{\mu\nu}u^\mu u^\nu,
\end{equation}
or: 
\begin{equation}
R_{\mu\nu}u^\mu u^\nu = -8\pi T_{\mu\nu}u^\mu u^\nu.
\end{equation}
Since the timelike unit vector field $u^\mu$ is arbitrary, we must have:
\begin{equation}
R_{\mu\nu} = -8\pi T_{\mu\nu},
\end{equation}
which is exactly Einstein's field equation 
\begin{equation}
R_{\mu\nu} = -8\pi(T_{\mu\nu} - \frac{1}{2}g_{\mu\nu}T^\alpha_\alpha),
\end{equation}
or
\begin{equation}
R_{\mu\nu} - \frac{1}{2}g_{\mu\nu}R = -8\pi T_{\mu\nu}
\end{equation}
in the special case, where Eq.(7.26) holds, i.e. the energy momentum stress tensor is
traceless. In other words, we have obtained Einstein's field equation in the special 
case, where matter consists of massless radiation in thermal equilibrium.

   Our derivation of Einstein's field equation was entirely based on Eq.(6.21), the 
fundamental equation of the thermodynamics of spacetime in our model. That equation
states just the conservation of energy, when the matter flowing thorugh an acceleration
surface consists of radiation, and the energy it carries of heat only. Actually, we did
not even need the result that the entropy of a spacelike two-surface of spacetime is, in
natural units, exactly one-half of its area in the low temperature limit. Our succesful 
derivation of Einstein's field equation provides a strong argument for the validity of
Eq.(6.21), as well as for the idea that one may meaningfully associate the concept
of heat with the acceleration surfaces. 

  It is instructive to consider the entropy exchange between the radiation and the 
accelerating plane. When the flow of heat carried by radiation through the plane increases
in the proper time, Eq.(6.21) implies that the plane shrinks. In the low temperature limit
the entropy loss of the plane is exactly one-half of the decrease in the area of the plane.
Eq.(6.26) implies that between the entropies $S_{rad}$ and $S_{plane}$ of the radiation 
and the plane, respectively, there is the following relationship at the point $P$:
\begin{equation}
\frac{d^2}{d\tau^2}(S_{rad} + S_{plane}) 
= (1 - \frac{T_{rad}}{T_U})\frac{d^2S_{rad}}{d\tau^2}.
\end{equation}
In this equation $T_{rad}$ is the temperature of the radiation, and $T_U$ is the Unruh
temperature measured by an observer at rest with respect to the plane. Hence we observe
that if $\frac{d^2S_{rad}}{d\tau^2}>0$, i.e. the flow of entropy carried by the radiation
through the plane increases in the proper time at the point $P$, then $T_{rad}$ must be
smaller than, or equal to, the Unruh temperature $T_U$, since otherwise the second law of
thermodynamics would be violated (It should be recalled that, at the point $P$, 
$\frac{dA}{d\tau}=0$, which implies that $\frac{dS_{plane}}{d\tau}=0$.). On the other hand, 
if $\frac{d^2S_{plane}}{d\tau^2}<0$, i.e. the flow of entropy carried by the radiation 
through the plane decreases in the proper time at the point $P$, we must have $T_{rad}$ 
greater than, or equal to, the Unruh temperature $T_U$. So we find that if we want to use
Eq.(6.21) in the derivation of Einstein's field equation, no matter whether the 
flow of entropy through the plane is increasing or decreasing in the proper time, we
must have
\begin{equation}
T_{rad} = T_U.
\end{equation}
In other words, we fix the proper acceleration $a$ of the plane such that the Unruh
temperature $T_U$ measured by an observer at rest with respect to the plane is exactly 
the same as is the absolute temperature $T_{rad}$ measured by that observer fo the 
radiation. In that case the plane and the radiation are in a thermal equilibrium with
each other, and the entropy of the plane is exactly converted to the entropy of the 
radiation, and vice versa. Actually, somewhat similar assumption was also used by Jacobson
in his derivation of Einstein's field equation by means of thermodynamical arguments. 
\cite{yksi}

   After obtaining Einstein's field equation, when matter consists of massless, 
non-interacting radiation in thermal equilibrium, the next challenge is to derive that equation
for general matter fields. In doing so, however, we meet with some difficulties, because
Eq.(6.21) is assumed to hold for {\it radiation} only. Moreover, the rate of change in the 
boost energy flow through the plane should be, in effect, the rate of change in the flow
of {\it heat}. The problem is that for general matter fields other forms of energy, except heat
(mass-energy, for instance) flow through the plane, and therefore it seems that we cannot
use the same kind of reasoning as we did above.

   To overcome these difficulties, let us assume that at the point $P$ our plane moves
with respect to the matter fields to the direction of the positive $z$-axis with a velocity
$v$, which is very close to the speed of light. In this limit the particles of the matter
fields move with enormous speeds through the plane to the direction of the negative 
$z$-axis, and their kinetic energies vastly exceed, in the rest frame of the plane, all the 
other forms of energy. This means that we may consider matter, in effect, as a gas of 
non-interacting massless particles, regardless of the kind of matter we happen to have.
In other words, the matter behaves, from the point of view of an observer at rest with
respect to the plane, as non-interacting, massless radiation. It was shown in Ref.[28], by
means of the first law of thermodynamics, that in the limit, where $v$ gets close to the
speed of light, and the proper acceleration $a$ is very large, the rate of change in the 
flow of the boost energy through the plane is exactly the rate of change in the flow of
heat through the plane. So we find that in the limit, where $v\rightarrow 1$, the speed
of light in the natural units, we are allowed to use Eq.(6.21) for general matter fields.

  If our plane moves, at the point $P$, with a velocity $v$ to the direction of the positive
$z$-axis and, at the same time, is in a uniformly accelerating motion with a constant proper
acceleration $a$ to the direction of the negative $z$-axis, the only non-zero components
of the vectors $u^\mu$ and $n^\mu$ are, in the orthonormal geodesic system of coordinates 
associated with the point $P$:
\begin{subequations}
\begin{eqnarray}
u^0 &=& \cosh\phi,\\
u^3 &=& \sinh\phi,\\
n^0 &=& \sinh\phi,\\
n^3 &=& \cosh\phi,
\end{eqnarray}
\end{subequations}
where
\begin{equation}
\phi := \sinh^{-1}(\frac{v}{\sqrt{1-v^2}})
\end{equation}  
is the boost angle. It is convenient to write $v$ in terms of a new parameter 
$\epsilon\in(0,1)$ such that
\begin{equation}
v = \frac{1-\epsilon}{1+\epsilon}.
\end{equation}   
As one may observe, the velocity $v$ goes to zero, when $\epsilon$ goes to one, and
it goes to one, the speed of light in the natural units, when $\epsilon$ goes to zero.
Using Eq.(7.38) we find:
\begin{subequations}
\begin{eqnarray}
u^\mu &=& \frac{1}{2}(\frac{k^\mu}{\sqrt{\epsilon}} + \sqrt{\epsilon}\,l^\mu),\\
n^\mu &=& \frac{1}{2}(\frac{k^\mu}{\sqrt{\epsilon}} - \sqrt{\epsilon}\,l^\mu),
\end{eqnarray}
\end{subequations}
where $k^\mu:=(1,0,0,1)$ and $l^\mu:=(1,0,0,-1)$ are future pointing null vectors.
This means that when the observer at rest with respect to our plane moves very close 
to his Rindler horizon, his world line seems to lie very close to the null geodesic
generated by the null vector $k^\mu$. When the proper acceleration $a$ goes to infinity,
the null vector $k^\mu$ becomes a generator of the past Rindler horizon of the observer,
whereas the null vector $l^\mu$ becomes a generator of the future Rindler horizon.

  Assuming that the plane moves, at the point $P$, with a very great speed with respect
to the matter fields, i.e. the components of the tensor $T_{\mu\nu}$ are fixed and finite
at the orthonormal geodesic system of coordinates associated with the point $P$, we find,
using Eqs.(7.23b) and (7.41), that the rate of change in the flow of heat through the plane
is:
\begin{equation}
\frac{\delta^2 Q}{d\tau^2}\vert_{\tau=0} = \frac{a}{2\epsilon}AT_{\mu\nu}k^\mu k^\nu 
+ O(\epsilon),
\end{equation}
where $O(\epsilon)$ denotes the terms, which are of the order $\epsilon$, or higher. It
was shown in Ref.\cite{kaksikymmentakuusi} that
\begin{equation}
\lim_{\epsilon\rightarrow 0}\vert w^3_{;3}\vert = 0.
\end{equation}
This means that if we consider a spatial region of spacetime in a frame of reference which
moves with a very large speed with respect to that region, the curvature of spacetime may
make the region to expand and contract in the directions orthogonal to the direction of the
movement of the region only. So we may use Eq.(7.18), when we consider the change of the
area $A$ of our plane. Eqs. (7.18) and (7.41) imply:
\begin{equation}
\frac{d^2A}{d\tau^2} = \frac{1}{4\epsilon}AR_{\mu\nu}k^\mu k^\nu + O(1),
\end{equation}
where $O(1)$ denotes the terms of the order $\epsilon^0$, or higher. It is obvious that
since $\epsilon$ is assumed to be very small, the leading terms in Eqs.(7.42) and (7.45)
are the ones, which are proportional to $\frac{1}{\epsilon}$.

   We may identify $\frac{\delta^2 Q}{d\tau^2}$ of Eq.(7.42) as 
$\frac{\delta^2 Q_{rad}}{d\tau^2}$
of Eq.(6.24). So we find that Eq.(6.24) implies, in the limit, where 
$\epsilon\rightarrow 0$: 
\begin{equation}
\frac{a}{2}AT_{\mu\nu}k^\mu k^\nu = -\frac{a}{16\pi}AR_{\mu\nu}k^\mu k^\nu,
\end{equation}
or: 
\begin{equation}
R_{\mu\nu}k^\mu k^\nu = -8\pi T_{\mu\nu} k^\mu k^\nu.
\end{equation}
Because $k^\mu$ may be chosen to be an arbitrary, future orientated null vector, we must 
have:
\begin{equation}
R_{\mu\nu} + fg_{\mu\nu} = -8\pi T_{\mu\nu},
\end{equation}
where $f$ is some function of the spacetime coordinates. It follows from the Bianchi 
identity 
\begin{equation}
(R^\mu_{\,\,\,\nu} - \frac{1}{2}R\delta^\mu_{\,\,\,\nu})_{;\mu} = 0,
\end{equation}
that 
\begin{equation}
f = -\frac{1}{2}R + \Lambda
\end{equation}
for some constant $\Lambda$, and hence we arrive at the equation
\begin{equation}
R_{\mu\nu} - \frac{1}{2}g_{\mu\nu} R + \Lambda g_{\mu\nu} = -8\pi T_{\mu\nu},
\end{equation}
which is Einstein's equation with a cosmological constant $\Lambda$. We have thus achieved
our goal: We have obtained Einstein's field equation from our model in the long distance
limit. 

    Our derivation of Einstein's field equation was based on the use of Eq.(6.21), the 
fundamental equation of the thermodynamics of spacetime in our model. However, we would
also have been able to derive that equation by assuming that when the boost energy flow 
through an accelerating plane is exactly the heat flow, and the temperature of the matter
flowing through the plane is exactly the Unruh temperature measured by an observer at rest
with respect to the plane, then the rate of change in the entropy flow through through the
plane is, in natural units, exactly one-half of the rate of change in the shrinking speed of
the plane. In other words, Einstein's equation may be viewed not only as a consequence of
Eq.(6.21), but also as a consequence of Eq.(1.2), and the result that every spacelike
two-surface of spacetime possesses entropy which, in the low temperature limit and in
natural units, is one-half of its area. 
In this sense our derivation of Einstein's field equation by means of 
thermodynamical considerations bears a lot of similarities with Jacobson's derivation of 
Einstein's equation about one decade ago. \cite{yksi} However, there are certain radical differences
between these two derivations: Jacobson considered the boost energy flow through a {\it
horizon} of spacetime, whereas we considered the boost energy flow through an accelerating,
spacelike two-plane. Horizons of spacetime are null hypersurfaces of spacetime, and 
therefore they are created, when all points of a spacelike two-surface move along certain
{\it null} curves of spacetime. In contrast, our spacelike two-plane was assumed to move in
spacetime with a speed less than that of light, and therefore all of its points move along
certain {\it timelike} curves of spacetime. Because of that our two-plane should not be 
considered as a part of any horizon of spacetime.

  Another important difference between the thermodynamical considerations of Jacobson
and ours is that Jacobson's approach leaves open the question about the precise numerical
value of the cosmological constant $\Lambda$, whereas our approach yields the result that
the cosmological constant $\Lambda$ must vanish:
\begin{equation}
\Lambda = 0.
\end{equation}
The vanishing of the cosmological constant is a direct consequence of Eq.(7.33), which 
describes the interaction between spacetime and massless radiation with traceless 
$T_{\mu\nu}$. That equation does not involve $\Lambda$, and therefore we may conclude  
that $\Lambda$ must vanish.

\maketitle

\section{The High Temperature Limit}

  So far we have considered the {\it low temperature limit} of our model. More precisely,
we have assumed that in the spacetime region under consideration most microscopic quantum 
black holes constituting spacetime are in their ground states. A mathematical expression for 
this assumption was that when we pick up a spacelike two-graph $\Sigma^{(2)}$ of spacetime,
then $N$, the number of microscopic black holes on $\Sigma^{(2)}$, is much bigger than 
$A/\alpha$, where $A$ is the area of $\Sigma^{(2)}$, and $\alpha=2\ln 2$. We have shown that
the Unruh and the Hawking effects, as well as Einstein's field equation, are straightforward
concequences of our model in the low temperature limit.

  The question is: What happens, if the temperature is {\it high} such that the assumption
$N>>A/\alpha$ is abandoned, and most microscopic black holes are in highly excited states? 
Presumably we shall encounter entirely new effects, which go beyond the classical and the 
semiclassical treatments of gravity.

   The object of this Section is to investigate these effects. The first task is to consider
the statistics of spacetime, when $A/\alpha>N$ on an arbitrary spacelike two-graph 
$\Sigma^{(2)}$ of spacetime. We shall keep all of the postulates of Section 4. Recall that
with those postulates the statistical weight $\Omega$ of a given macrostate of 
$\Sigma^{(2)}$ is the number of strings $(n_1,n_2,...,n_m)$, where $1\leq m\leq N$, and 
$n_1,n_2,...,n_m\in\lbrace 1,2,3,...\rbrace$ such that $n_1+n_2+...+n_m=A/\alpha$. It is 
relatively easy to show that if we denote the positive integer $A/\alpha$ by $n$, then 
\begin{equation}
\Omega = \sum_{k=0}^{N-1}
\left(\begin{array}{ccc}
n-1\\
k\end{array} \right) 
\end{equation}
provided that $N<n$. If $N=n$, we get $\Omega=2^{n-1}$, and so we arrive, again, at 
Eq.(4.6), as expected. For instance, if $N=2$ and $n=3$, Eq.(8.1) implies that 
$\Omega=1+2=3$. Indeed, there are 3 ways to express 3 as a sum of at most 2 positive 
integers. More precisely, we have $3=2+1=1+2$.

  The numbers $n$ and $N$ are expected to be very large, presumably of order $10^{70}$. When
$n$ and $N$ are that large, it is very clumsy to operate with the expressions like (8.1) for
$\Omega$. To make our analysis more flexible, we should find an appropriate approximation
for $\Omega$, when both $n$ and $N$ are large. It is a well known result of the elementary
theory of probability that the Gaussian normal distribution function may be used to 
approximate the binomial distribution function such that \cite{kaksikymmentaseitseman}:
\begin{equation}
\left(\begin{array}{ccc}
n\\
x\end{array} \right)
p^x q^{n-x} \approx \frac{1}{\sqrt{2\pi npq}}\,e^{-\frac{(x-np)^2}{2npq}}
\end{equation}
for large $n$ and $x$. Setting $p=q=1/2$, we get:
\begin{equation}
\left(\begin{array}{ccc}
n\\
x\end{array} \right) \approx 2^n\sqrt{\frac{2}{n\pi}}\,e^{-\frac{2}{n}(x-\frac{n}{2})^2}.
\end{equation}
This expression gives an excellent approximation for the binomial coefficient as far as both
$x$ and $n$ are large, and $n$ is not more than five times bigger than $x$. When $x$ is less 
than one-fifth of $n$, the relative errors between the left and the right hand sides of 
Eq.(8.3) become intolerably large, and more accurate approximations for the binomial 
coefficient must be sought. 

   When both $n$ and $N$ are large in Eq.(8.1), we may replace the sum by an integral. Using
Eq.(8.3) we get:
\begin{equation}
\Omega \approx 2^{n-1}\sqrt{\frac{2}{(n-1)\pi}}\int_{x=0}^{N-1}\exp\lbrack 
-\frac{2}{n-1}(x-\frac{n-1}{2})^2\rbrack\,dx,
\end{equation}
which gives:
\begin{equation}
\Omega \approx 2^{n-2}\lbrace erf(\sqrt{\frac{n-1}{2}}) + 
erf\lbrack(N-1)\sqrt{\frac{2}{n-1}}-\sqrt{\frac{n-1}{2}}\rbrack\rbrace,
\end{equation}
where
\begin{equation}
erf(x) := \frac{2}{\sqrt{\pi}}\int_0^x e^{-u^2}\,du
\end{equation}
is the error function. Again, one finds that the right hand side of Eq.(8.5) gives a fairly
accurate approximation for $\Omega$, when $N$ is not less than about one-quarter of $n$. 
When $N=n$, we get:
\begin{equation}
\Omega \approx 2^{n-1} erf(\sqrt{\frac{n-1}{2}}),
\end{equation}
and since $\lim_{x\rightarrow\infty} erf(x)=1$, we observe that for large $n$, 
$\Omega=2^{n-1}$ with a very high precision.

   The entropy $S$ associated with the spacelike two-graph $\Sigma^{(2)}$ is the natural 
logarithm of $\Omega$. We get:
\begin{equation}
S = \frac{1}{2}A + \ln\lbrace\frac{1}{2} erf(\frac{1}{2}\sqrt{\frac{A}{\ln 2}}) 
+ \frac{1}{2} erf\lbrack 2N\sqrt{\frac{\ln 2}{A}} 
- \frac{1}{2}\sqrt{\frac{A}{\ln 2}}\rbrack\rbrace.
\end{equation}
When obtaining Eq.(8.8) we have used Eqs.(4.4) and (6.32). We have also employed the 
assumptions that $n$ and $N$ are large by replacing $n-1$ by $n$, and $N-1$ by $N$. The 
first term on the right hand side of Eq.(8.8) is the familiar expression for the entropy
of $\Sigma^{(2)}$ in the low temperature limit. Since the values of the error function lie
within the interval $[-1,1]$, we find that, when compared with the first term, the second
term is negligible, except when $N/\sqrt{A}$ is very small. In that limit, however, our
approximation is not valid.

   Eq.(8.8) gives an excellent approximation for the entropy $S$, when $N$ is close to $n$. 
It may therefore be used for an investigation of the stability of the macroscopic regions
of classical spacetime against small fluctuations in the average distributions of the area 
eigenvalues between the microscopic black holes. In these kind of fluctuations $n$ is only 
slightly bigger than $N$. We shall denote the difference between $n$ and $N$ by
\begin{equation}
\Delta N := n - N,
\end{equation}
and we shall study what happens, when $\Delta N/\sqrt{A}$ is much less than unity. We get:
\begin{equation}
S = \frac{1}{2}A + \ln\lbrack \frac{1}{2} erf(\frac{1}{2}\sqrt{\frac{A}{\ln2}} 
+ \frac{1}{2} erf(\frac{1}{2}\sqrt{\frac{A}{\ln2}} - 2\Delta N\sqrt{\frac{\ln 2}{A}})\rbrack.
\end{equation}
It follows from Eq.(8.6), the definition of the error function, that
\begin{equation}
D\,erf(x) = \frac{2}{\sqrt{\pi}}\,e^{-x^2},
\end{equation}
and therefore we may write for small $\Delta x$, as an excellent approximation:
\begin{equation}
erf(x+\Delta x) \approx erf(x) + \frac{2\Delta x}{\sqrt{\pi}}\,e^{-x^2}.
\end{equation}
Since the error function $erf(x)$ is a very slowly varying function of $x$ for large $x$, 
Eq.(8.12) holds for large $x$ even when $\Delta x$ is not very small. Applying that equation
for Eq.(8.10) we get:
\begin{equation}
S \approx \frac{1}{2} A + \ln\lbrack erf(\frac{1}{2}\sqrt{\frac{A}{\ln 2}}) 
- \frac{2\Delta N}{\sqrt{\pi}}\sqrt{\frac{\ln 2}{A}}\exp(-\frac{1}{4}\frac{A}{\ln2})\rbrack.
\end{equation}
Since $A$ is assumed to be very large, we may approximate 
$erf(\frac{1}{2}\sqrt{\frac{A}{\ln 2}})$ by one. So we finally have:
\begin{equation}
S \approx \frac{1}{2} A - \frac{2\Delta N}{\sqrt{\pi}}\sqrt{\frac{\ln2}{A}}
\exp(-\frac{1}{4}\frac{A}{\ln 2}).
\end{equation}
When $A$ is increased, the second term on the right hand side goes very rapidly to zero, and
its contribution to the entropy of the spacelike two-graph $\Sigma^{(2)}$ is negligible.
So we find that macroscopic regions of spacetime are indeed very stable against the small
fluctuations in the quantum states of the microscopic quantum black holes.

  Quite another matter is what happens when, in a sufficiently small region of spacetime, 
the fluctuations in the quantum states are very large, i.e. when $n>>N$. In that case we 
shall need, instead of Eq.(8.3), an approximation for the binomial coefficient 
$\left(\begin{array}{ccc}
n\\
k\end{array} \right)$, when $n$
is very much larger than $k$. To this end, let us write the binomial coefficient as a 
product:
\begin{equation}
\left(\begin{array}{ccc}
n\\
k\end{array} \right) = \frac{1}{k!}n(n-1)(n-2)\cdots(n-k+1).
\end{equation}
It is easy to see that if $n/N>>1$, then
\begin{equation}
\left(\begin{array}{ccc}
n\\
k\end{array} \right) \sim \frac{1}{k!} n^k.
\end{equation}
In other words, if the left hand side of Eq.(8.16) is divided by the right hand side, then 
the resulting quotient goes to unity in the limit, where $n/k\rightarrow\infty$. The 
asymptotic approximation given by Eq.(8.18) is tolerably precise. For instance, for the binomial 
coefficient 
$\left(\begin{array}{ccc}
200\\
5\end{array} \right)$ the relative error made in the approximation is less than 5 per 
cents. Writing the binomial coefficients on the right hand side of Eq.(8.1) as products, and
taking the sum one finds that the last term of the sum dominates. Hence we may write:
\begin{equation}
\Omega = \sum_{k=0}^{N-1}\left(\begin{array}{ccc}
n-1\\
k\end{array} \right) \sim \frac{1}{(N-1)!}(n-1)^{N-1}.
\end{equation}
This approximation, also, is fairly precise: For instance, when $n=200$ and $N=7$, the 
relative error made in the approximation is less than 5 per cents.

  In the limit $n/N>>1$ the entropy of the spacelike two-graph $\Sigma^{(2)}$ is now:
\begin{equation}
S = \ln\Omega \approx (N-1)\ln(n-1) - (N-1)\ln(N-1) + N-1,
\end{equation}
where we have used Stirling's formula. The last two terms on the right hand side of 
Eq.(8.18) are mere additive constants with respect to the area $A$ of $\Sigma^{(2)}$, and 
they may therefore be neglected. ($N$ represents here a particle number, and all physically relevant
quantities obtainable from the entropy $S$ of a system are those partial derivatives of $S$,
where $N$ has been kept as a constant during the differentiation.). When we write $S$ in 
terms of $A$, and abandon the physically irrelevant additive constants, we get:
\begin{equation}
S = N\ln A.
\end{equation}
In other words, when the microscopic quantum black holes in the spacetime region under 
consideration are, in average, in highly excited states, the entropy $S$ is not proportional
to the area $A$, but it depends {\it logarithmically} on $A$. It has been speculated for a 
long time by several authors that there might be, in addition to a simple proportionality,
a logarithmic dependence between area and entropy \cite{kaksikymmentakahdeksan}.
Our model implies that there indeed 
exists such a dependence, and this logarithmic dependence dominates in a certain limit.
It should be noted that when obtaining Eq.(8.19) we took into account that $n$ and $N$ are 
both very large, and therefore we were, again, justified to replace $n-1$ by $n$, and 
$N-1$ by $N$.

  Consider now the physical consequences of Eq.(8.19). We find that between the 
infinitesimal changes in the entropy $S$ and the area $A$ there is a relationship:
\begin{equation}
dS = \frac{N}{A}\,dA.
\end{equation}
As it was discussed in Section 6, the area of a given spacelike two-surface of spacetime
may change, for instance, when radiation goes through that two-surface. It follows from 
the results and the definitions of Section 6 that if we have an acceleration surface, then 
the change in the {\it heat content} of the surface is
\begin{equation}
\delta Q = \frac{a}{4\pi}\,dA,
\end{equation}
provided that $a$ is kept as a constant during the variation of $Q$. Because, according to 
the first law of thermodynamics,
\begin{equation}
\delta Q = T\,dS,
\end{equation}
Eqs.(8.20) and (8.21) imply:
\begin{equation}
NT = \frac{a}{4\pi}A.
\end{equation}
Using Eq.(8.21) we get:
\begin{equation}
\delta Q = N\,dT
\end{equation}
or, in SI units:
\begin{equation}
\delta Q = Nk_B\,dT.
\end{equation}

   To some extent, Eq.(8.25) may be used as a consistency check of our model. The limit
$n/N>>1$ corresponds to the {\it high temperature limit}, where most microscopic quantum 
black holes on $\Sigma^{(2)}$ lie on the highly excited states. In this limit the thermal 
fluctuations of the areas of the black holes on $\Sigma^{(2)}$ become so large that the
quantum mechanical discreteness of the area spectrum is, in effect, washed out into 
continuum, and classical statistical mechanics may be applied. It is a general feature of
almost any system that in a sufficiently high temperature the relationship between the 
infinitesimal changes in the heat $Q$, and in the absolute temperature $T$ of
the system is of the form:
\begin{equation}
\delta Q = \gamma Nk_B\,dT.
\end{equation}
In this equation, $N$ is the number of the constituents (atoms or molecules) of the system,
and $\gamma$ is a pure number of order one, which depends on the physical properties (the 
number of independent degrees of freedom, etc.) of the system. For instance, in sufficiently
high temperatures most solids obey the Dulong-Petit law \cite{kaksikymmentayhdeksan}:
\begin{equation}
\delta Q = 3Nk_B\,dT,
\end{equation}
where $N$ is the number of molecules in the solid. From Eq.(8.25) we observe that this 
general feature is also possessed by our spacetime model. Eq.(8.25) is exactly what one
expects in the high temperature limit on grounds of general statistical arguments.

   According to Eq.(8.23) the absolute temperature $T$ measured by an observer at rest
with respect to an accelerating, spacelike two-surface depends, in the high temperature 
limit, on the proper acceleration $a$, area $A$, and the number $N$ of the microscopic 
quantum black holes on the surface such that:
\begin{equation}
T = \frac{a}{4\pi N}A
\end{equation}
or, in SI units:
\begin{equation}
T = \frac{c^2 a}{4\pi Nk_BG}A.
\end{equation}
Our derivation of Eq.(8.29) was analogous to the derivation of the exression for the 
Unruh temperature $T_U$ in Eq.(6.32). Indeed, Eq.(8.29) represents, in the high temperature
limit, where $a$ becomes very large, the minimum temperature an accelerating observer may 
measure. In other words, Eq.(8.29) gives the Unruh temperature measured by an accelerating
observer in the large $a$ limit. As one may observe, Eq.(8.29) is radically different from
Eq.(6.32): In the large $a$ limit the Unruh temperature is not constant for constant $a$, 
but it is directly proportional to the ratio $A/N$.

  One of the most intriguing problems of semiclassical gravity, which considers spacetime 
classically, and matter fields quantum mechanically, concerns the behavior of black holes 
at the last stages of their evaporation: According to Eq.(6.38) the Hawking temperature of 
an evaporating black hole increases, while its mass decreases. Will the black hole finally
vanish in a cataclysmic explosion, or shall we be left with a small remnant of the hole?

  It appears that this question may probably be answered within the context of our model of 
spacetime: If we substitute for $A$ $4\pi r^2$, which gives the area of a two-sphere 
surrounding the event horizon of a Schwarzschild black hole, and for $a$ the expression
the expression for $a^\mu n_\mu$ in Eq.(6.46), we find that in the high temperature limit
the Hawking temperature of a Schwarzschild black hole measured by an observer just outside
of the event horizon of the hole is:
\begin{equation}
T_H = (1 - \frac{2M}{r})^{-1/2}\frac{M}{N}
\end{equation}
or, in SI units:
\begin{equation}
T_H = (1 - \frac{2GM}{c^2 r})^{-1/2}\frac{Mc^2}{Nk_B}.
\end{equation}
As one may observe, for small enough black holes the Hawking temperature will no more be
inversely, but {\it directly} proportional to the Schwarzschild mass $M$ of the hole. This
means that at a certain stage the evaporation will cease, and a tiny remnant of the hole
will be left.

   It is interesting that neither of the Eqs.(8.29) and (8.31) involve the Planck constant
$\hbar$. This is something one might expect: In very high temperatures the thermal 
fluctuations in the horizon area eigenvalues of the microscopic quantum black holes become
so large that the discrete area spectrum predicted by quantum mechanics is, in effect, 
washed out into continuum. In this limit quantum effects on the statistics of spacetime
may be ignored, and classical statistics may be applied. In contrast to the high temperature
limit, where quantum effects become negligible, in the {\it low temperature limit} the
quantum effects of spacetime play an essential role. In this sense we may say that contrary
to the common beliefs, spacetime behaves quantum mechanically in low temperatures, and 
classically in high temperatures. This explains the presence of the Planck constant 
$\hbar$ in the low temperature formulas (6.32) and (6.39), and its absence in the high 
temperature formulas (8.29) and (8.31): Eqs.(6.32) and (6.39) are quantum mechanical,
whereas Eqs.(8.29) and (8.31) are classical. 

  Which form will Einstein's field equation take in the high temperature limit? To answer 
this question, recall that in Section 7 we obtained Einstein's field equation by means of 
our fundamental thermodynamical equation (6.21) only. Nowhere in our derivation did we 
use any explicit relationship between the area and the entropy of a spacelike two-surface
of spacetime. Since the temperature of spacetime has effects on this relationship only,
we find that if we assume that Eq.(6.21) holds as such for all temperatures, no matter
whether those temperatures are low or high, Einstein's field equation is independent of 
that temperature. In other words, Einstein's field equation takes in high temperatures 
exactly the same form as it does in low temperatures. Since Einstein's field equation
is independent of the relationship between area and entropy, it is independent of the 
precise microscopic physics of spacetime. This startling conclusion reminds us of Nielsen's
famous idea of Random Dynamics \cite{kymmenen}. According to Nielsen's idea nature behaves 
in such a 
way that no matter what we assume of its behavior at the Planck energy scales, the 
consequences of those assumptions will always produce, as sort of statistical averages, the
well known laws of physics in the low energy limit. Indeed, our model is in harmony with 
this idea: The high energy effects such as the Unruh and the Hawking effects, depend 
crucially on the microphysics of spacetime, whereas the low energy effects, such as 
classical gravity, are independent of that microphysics.

\maketitle

\section{A Quantum Theory of Microscopic Black Holes}

 The starting point of our quantum mechanical model of spacetime was an assumption that
spacetime consists of microscopic quantum black holes with a horizon area spectrum given by
Eq.(2.3). Of course, it would be possible to take Eq.(2.3) just as one of the postulates of 
our model, and then work out the consequences. Actually, this is exactly what we have done
in this paper. Our model would not be complete, however, and probably not even plausible,
without a reasonably well defined quantum theory of microscpic black holes, based on the 
basic principles of quantum mechanics, and with a property of producing Eq.(2.3). 
Fortunately, such a theory already exists, and it was published in Ref.\cite{yksitoista}. For the sake of 
completeness, it is appropriate to include a brief review and a critical analysis of that 
theory here.

  The quantum theory in question is based on the canonical quantization of spherically 
symmetric, asymptotically flat vacuum spacetimes. According to Birkhoff's theorem the only
spherically symmetric, asymptotically flat vacuum solution to Einstein's field equation
is the Schwarzschild solution, and therefore the theory may be viewed as a canonical quantum
theory of Schwarzschild black holes. The starting point of the theory is Kuchar's analysis
of the classical Hamiltonian dynamics of spherically symmetric, asymptotically flat vacuum 
spacetimes \cite{kolmekymmenta}. Kuchar wrote the spacetime metric of such spacetime in the Arnowitt-Deser-Misner
(ADM) form:
\begin{equation}
ds^2 = -N^2\,dt^2 + \Lambda^2(dr + N^r\,dt)^2 + R^2\,d\Omega^2,
\end{equation}
where $d\Omega^2$ is the metric on the unit two-sphere, and the lapse $N$, the shift $N^r$,
as well as the dynamical variables $\Lambda$ and $R$ of the spacetime geometry are assumed 
to be functions of the time coordinate $t$, and of the radial coordinate $r$ only. The 
spacelike hypersurfaces, where $t=constant$ are assumed to cross the Kruskal spacetime from
the left to the right asymptotic infinities in arbitrary ways.

  After writing the spacetime metric in the form given by Eq.(9.1), Kuchar proceeded to 
write the Einstein-Hilbert action $S$, supplemented with appropriate boundary terms at 
asymptotic infinities, in the Hamiltonian form:
\begin{equation}
S = \int dt\int_{-\infty}^{\infty} dr(P_\Lambda\dot{\Lambda} + P_R\dot{R} - N{\cal H} 
- N^r{\cal H}_r) - \int dt(N_+(t)M_+(t) + N_-(t)M_-(t)).
\end{equation}
In this equation the dot means time derivative, and $P_\Lambda$ and $P_R$, respectively, are
the canonical momenta conjugate to $\Lambda$ and $R$. ${\cal H}$ and ${\cal H}_r$, 
respectively, are the Hamiltonian and the momentum constraints, and $M_+(t)$ and $M_-(t)$, 
respectively, are the ADM energies at the right and the left asymptotic infinities. 
$N_+(t)$ and $N_+(t)$ are the lapse functions at those infinities. (For the detailed 
expressions of the constraints ${\cal H}$ and ${\cal H}_r$ as well as the fall-off
conditions of the dynamical variables, and their conjugate momenta, see Ref.
\cite{kolmekymmenta}.).

  As the next step Kuchar performed a canonical transformation from the variables $\Lambda$,
$R$, $P_\Lambda$ and $P_R$ to certain new canonical variables $M$, $R$, $P_M$ and 
$\tilde{P}_R$, and he wrote the Hamiltonian form of the action in terms of these new 
variables as:
\begin{equation}
S = \int dt\int_{-\infty}^{\infty} dr(P_M\dot{M} + \tilde{P}_R\dot{R} - N{\cal H} 
- N^r{\cal H}_r) - \int dt(N_+(t)M_+(t) + N_-(t)M_-(t)). 
\end{equation}
An advantage of the new canonical coordinates is that the action simplifies considerably, 
when the constraint equations
\begin{subequations}
\begin{eqnarray}
{\cal H} &=& 0,\\
{\cal H}_r &=& 0,
\end{eqnarray}
\end{subequations}
are taken into account: It turns out that the constraint equations imply:
\begin{subequations}
\begin{eqnarray}
\frac{\partial M}{\partial r} &=& 0,\\
\tilde{P}_R &=& 0.
\end{eqnarray}
\end{subequations}
In other words, $M(r,t)$ is actually a function of the time coordinate $t$ only, i.e. we
may write:
\begin{equation}
M(r,t) = m(t),
\end{equation}
where $m(t)$ is some function of time, and $\tilde{P}_R$ vanishes. Moreover, it turns out 
that the falloff conditions for the variables $\Lambda$, $R$, $P_\Lambda$ and $P_R$ imply
that $M(r,t)$ coincides with $M_\pm(t)$ at asymptotic inifinities. Hence it follows that 
the action may be written in the reduced physical phase space, when the constraint equations
are satisfied, in a very simple form:
\begin{equation}
S = \int dt\lbrack p_m\dot{m} - (N_+ + N_-)m\rbrack,
\end{equation}
where we have defined:
\begin{equation}
p_m(t) := \int_{-\infty}^\infty P_M(t,r)\,dr.
\end{equation}
In other words, the action may be written in terms of variables which no more depend on $r$,
but on the time $t$ only. This means that the classical Hamiltonian theory of spherically 
symmetric, asymptotically flat vacuum spacetimes is no more a field theory, but a theory 
with a finite number of degrees of freedom.

  The variable $p_m$ is the canonical momentum conjugate to $m$. In terms of the canonical
coordinates $m$ and $p_m$ the classical Hamiltonian takes the form:
\begin{equation}
H = (N_+ + N_-)m.
\end{equation}
The Hamiltonian equations of motion for the variables $m$ and $p_m$ are:
\begin{subequations}
\begin{eqnarray}
\dot{m} &=& \frac{\partial H}{\partial p_m} = 0,\\
\dot{p}_m &=& -\frac{\partial H}{\partial m} = -(N_+ + N_-).
\end{eqnarray}
\end{subequations}
These equations imply that $m$ is a constant and may be identified as the Schwarzschild
mass $M$ of the Schwarzschild black hole. Moreover, $p_m$ gives, up to an additive constant,
the negative of the sums of the elapsed proper times at asymptotic infinities.

  The lapse functions $N_+$ and $N_-$ determine the gauge of the theory. More precisely,
they determine the choice of the time coordinate at asymptotic infinities. From the physical
point of view it is sensible to choose the gauge, where
\begin{subequations}
\begin{eqnarray}
N_+ &\equiv& 1,\\
N_- &\equiv& 0,
\end{eqnarray}
\end{subequations}
In other words, at the right infinity we have chosen the time coordinate to be the 
asymptotic Minkowski time, and we have "frozen" the time evolution at the left infinity. In
this gauge the classical Hamiltonian is simply:
\begin{equation}
H = m.
\end{equation}
This choice of gauge is sensible, if we consider a Schwarzschild black hole from the point 
of view of an asymptotic observer at the right infinity, at rest with respect to the hole:
Such an observer makes observations at just one of the infinities, the ADM energy of 
spacetime he observes is just the Schwarzschild mass of the hole, and his time coordinate 
is the asymptotic Minkowski time.

  Although the canonical coordinates $m$ and $p_m$ are very simple conceptually, there is
a very grave disadvantage in them: They describe the {\it static} aspects of spacetime 
containing a Schwarzschild black hole only. However, there is {\it dynamics} in such 
spacetimes, in the sense that in the spacetime region, where the Schwarzschild coordinate
$r<2M$, it is impossible to choose a timelike coordinate such that the geometry of 
spacetime with respect to this time coordinate would be static. To describe the dynamical aspects
of Schwarzschild black hole spacetimes it is useful to perform a canonical transformation 
from the canonical coordinates $m$ and $p_m$ to the new canonical coordinates $a$ and $p_a$
such that, when written in terms of these coordinates, the classical Hamiltonian of 
Eq.(9.12) takes the form (for the details of this canonical transformation, see Ref.\cite{yksitoista}):
\begin{equation}
H = \frac{1}{2a}(p_a^2 + a^2).
\end{equation}

   It is easy to see what is the geometrical interpretation of the canonical coordinates 
$a$ and $p_a$: If we write the Hamiltonian equation of motion for $p_a$:
\begin{equation}
\dot{a} = \frac{\partial H}{\partial p_a} = \frac{p_a}{a},
\end{equation}
and use Eq.(9.12), we get:
\begin{equation}
\dot{a} = \pm\sqrt{\frac{2M}{a}-1},
\end{equation}
where we have identified $m$ with the Schwarzschild mass $M$. This equation should be 
compared with an equation of motion of an observer in a radial free fall through the bifurcation
two-sphere in the Kruskal spacetime:
\begin{equation}
\dot{r} = \pm\sqrt{\frac{2M}{r} - 1},
\end{equation}
where the dot means proper time derivative. As one may observe, Eqs.(9.15) and (9.16) are 
identical: In Eq.(9.15) we have just replaced the Schwarzschild coordinate $r$ of Eq.(9.16)
by the canonical coordinate $a$. Hence we may obtain the geometrical interpretation of $a$:
It gives the radius of the Einstein-Rosen wormhole throat in a foliation, where the time 
coordinate is chosen to be the proper time of an asymptotic observer in a free fall through
the bifurcation two-sphere in the Kruskal spacetime. 

  When we chose the lapse functions $N_\pm$ at asymptotic infinities as in Eq.(9.11), we 
picked up a gauge, where the time coordinate at the right infinity coincides with the 
asymptotic Minkowski time of a faraway observer. We should therefore find a foliation of 
spacetime, where the time coordinate of the faraway observer coincides with the proper time
of the freely falling observer at the throat. These kind of foliations clearly exist. A 
concrete example is provided by the Novikov coordinates (for the detailed properties of 
these coordinates see Ref.\cite{kuusi}.). It should be noted, however, that all these foliations
are incomplete in a sense that they are valid during a finite time interval only: For
the observer in a free fall at the wormhole throat it takes the proper time
\begin{equation}
\tau = 2\int_0^{2M}\frac{1}{\sqrt{\frac{2M}{r} - 1}}\,dr = 2\pi M
\end{equation}
to fall from the past to the future singularity in the Kruskal spacetime. Hence it follows 
that our foliation is valid only when the time coordinate $t$ lies within the interval
$\lbrack -\pi M, \pi M\rbrack$. We shall see in a moment, however, how this classical 
incompleteness is removed by quantum mechanics.

  The transition from the classical to the quantum theory of spherically symmetric, 
asymptotically flat vacuum spacetimes is straightforward: We define the Hilbert space to be
the space $L^2(\Re^+,a^s\,da)$, where $s$ is an arbitrary real, and the inner product is 
defined as:
\begin{equation}
\langle\psi_1\vert\psi_2\rangle := \int_0^\infty \psi^*_1(a)\psi_2(a)a^s\,da,
\end{equation}
for every function $\psi_{1,2}(a)\in L^2(\Re^+,a^s\,da)$. The elements $\psi(a)$ of the 
Hilbert space are the {\it wave functions} of the Schwarzschild black hole, and they are 
functions of the throat radius $a$. The canonical momentum $p_a$ conjugate to $a$ is 
replaced by the operator
\begin{equation}
\hat{p}_a := -i\frac{d}{da},
\end{equation}
and the classical Hamiltonian $H$ of Eq.(9.13) is replaced by the corresponding Hamiltonian
operator
\begin{equation}
\hat{H} := -\frac{1}{2}a^{-s}\frac{d}{da}(a^{s-1}\frac{d}{da}) + \frac{1}{2}a,
\end{equation}
which is symmetric with respect to the inner product (9.18). The eigenvalue equation for 
$\hat{H}$ reads as:
\begin{equation}
\left\lbrack -\frac{1}{2}a^{-s}\frac{d}{da}(a^{s-1}\frac{d}{da}) 
+ \frac{1}{2}a\right\rbrack\psi(a) 
= M\psi(a).
\end{equation}
This equation may be viewed as a sort of "Schr\"odinger equation" of the Schwarzschild 
black hole. The spectrum of $\hat{H}$ gives the (Schwarzschild) mass eigenvalues $M$ of the 
hole.

  The properties of Eq.(9.21) have been considered in details in Refs.\cite{yksitoista}
 and \cite{kahdeksan}. It turns 
out that no matter how we choose $s$ and the self-adjoint extension, the mass spectrum is 
always discrete and bounded from below. Moreover, there is always an extension, where the 
mass  spectrum is positive. Of particular interest is the case, where $s=2$. This choice
for $s$ might appear natural, because at the horizon, where $r=2M$, the variable $a$ 
coincides with the radial coordinate $r$. In that case one may show that the WKB estimate
for the large mass eigenvalues is of the form:
\begin{equation}
M^{WKB}_n = \sqrt{2n +1}\,M_{PL},
\end{equation}
where $n=0,1,2,...$, and $M_{Pl}$ is the Planck mass of Eq.(2.7). As one may observe, our 
WKB mass spectrum is identical to the mass spectrum of Eq.(2.6). Actually, a numerical 
analysis of Eq.(9.21) shows that the WKB estimate of Eq.(9.22) gives an excellent 
approximation for the mass eigenvalues even for small $n$: At the ground state, where $n=0$,
the difference between the computed mass eigenvalue, and our WKB estimate is less than 1 per cent, 
and the difference very rapidly goes to zero, when $n$ increases. So we may really 
construct a quantum mechanical model of a microscopic black hole, which produces, at least
as an excellent approximation, the mass spectrum of Eq.(2.6), and thereby the horizon area 
spectrum of Eq.(2.3).

   As it was noted before, our classical Hamiltonian theory of Schwarzschild black hole
spacetimes was incomplete in the sense that the spacelike hypersurfaces, where $t=constant$, 
do not cover the whole spacetime, but they run into a black hole singularity within a finite 
time. This classical incompleteness is very nicely removed by quantum mechanics: The wave 
packet corresponding to the throat radius $a$ is reflected from the singularity, and in a 
stationary state there is a standing wave between the event horizon and the singularity. So
there are no propagating wave packets between the horizon and the singularity, when the
system is in a stationary state, and our quantum theory is therefore valid in any moment of 
time.

  Actually, there is an interesting analogue between Eq.(9.21) and the Schr\"odinger 
equation of a hydrogen atom: When the hydrogen atom is in an s-state, where the orbital
angular momentum of the electron orbiting around the proton vanishes, the electron should,
classically, collide with the proton in a very short time. Quantum mechanically, however, 
the wave packet representing the electron is scattered from the proton, and finally the
electron is represented by a standing wave, which makes the quantum theory of the hydrogen
atom valid in any moment of time. In a Schwarzschild black hole, the proton is replaced by
the singularity, and the distance of the electron from the proton is replaced by the throat
radius $a$ of the hole. Nevertheless, the solution provided by quantum theory to the 
problems of the classical one are similar both in black holes and hydrogen atoms.

   Our quantum theory of microscopic black holes was based on a quantization of classical 
black hole spacetimes. In other words, we proceeded from classical general relativity to 
quantum black holes. The fundamental idea behind our quantum mechanical model of spacetime,
however, has been to do things in another way round: We have {\it assumed} that at the 
Planck length scale spacetime possesses a certain quantum mechanical microstructure, and 
from this microstructure we have obtained classical general relativity at macroscopic 
length scales. One may therefore wonder, what exactly is the role played by our quantum 
theory of microscopic black holes in our quantum mechanical model of spacetime.

  One possibility is to elevate Eq.(9.21) with $s=2$ to the status of one of the 
independent postulates of our model. Using that equation one may obtain the area spectrum 
of the microscopic quantum black holes, which constitute spacetime. As we saw, Eq.(9.21) 
produces the area spectrum of Eq.(2.3) with an excellent accuracy. If we adopt this point 
of view on the status of Eq.(9.21), we may consider the classical Hamiltonian dynamics 
of black hole spacetimes producing Eq.(9.21) when the standard principles of quantum 
mechanics are applied, as a mere source of inspiration and a heuristic aid of thought: The 
real physics of microscopic quantum black holes lies in Eq.(9.21), and the derivation of
Eq.(9.21) may be forgotten.

    Another point of view on the  status of Eq.(9.21) is provided by the black hole 
uniqueness theorems, which imply that after a black hole has settled down, a non-rotating,
uncharged black hole has just one physical degree of freedom, which may be taken to be 
the Schwarzschild mass or, equivalently, the event horizon area of the black hole. In this 
paper we have assumed that this same degree of freedom characterizes the properties of 
black holes down to the Planck scale. In other words, no matter whether a black hole is 
microscopic or macroscopic, its physical degree of freedom is the same. In this sense 
macroscopic black holes provide, in our model, a sort of "microscope" for the investigation
of gravity at the Planck length scale: Macroscopic black holes are similar to the 
microscopic ones--only their size is bigger. Since classical general relativity is the
long distance limit of our model, one may expect that quantization of classical black
hole spacetimes using classical general relativity as a starting point yields a correct
eigenvalue equation for the event horizon area even for microscopic black holes.

  \maketitle

\section{Concluding Remarks}

   In this paper we have considered a quantum mechanical model of spacetime, where spacetime
was constructed from Planck size black holes. Spacetime was assumed to be a graph, where
black holes lie on the vertices. The Planck size black holes constituting spacetime were 
assumed to be quantum mechanical objects with an equal spacing in their horizon area 
spectrum. In the low temperature limit, where most holes lie in their ground states, our
model implied that every spacelike two-surface of spacetime may be associated with an 
entropy, which is one-half of the area of that two-surface, whereas in the high 
temperature limit, where most black holes are in highly excited states, the entropy depends
logarithmically on the area. At macroscopic length scales our model implied Einstein's field
equation with a vanishing cosmological constant as a sort of thermodynamical equation of 
state of spaetime and matter fields. Our model also implied the Hawking and the Unruh 
effects in the low temperature limit. Moreover, our model implied that the area of any 
spacelike two-surface has an equal spacing in its spectrum such that its possible 
eigenvalues are of the form:
\begin{equation}
A_n = 2n\ln 2\,\ell_{Pl}^2,
\end{equation}
where $n$ is an integer, and $\ell_{Pl}$ is the Planck length.

   It is surprising how very simple assumptions about the structure of spacetime at the Planck
length scale imply, at macroscopic length scales, all of the "hard facts" of gravitational 
physics as such as we know them today. Indeed, after we were able to show that every 
spacelike two-graph may be associated with an entropy which is, in the low temperature 
limit, one-half of its area, everything else followed in a very simple and straightforward
manner. The key point of this paper was to abandon completely the idea that quantization of
gravity would involve, in one way or another, quantization of Einstein's classical general
relativity as if it were an ordinary field theory. It is possible that the very reason for
the physicists' failure to quantize gravity so far has been their natural prejudice that
general relativity would be a field theory in the same sense as, for instance, classical 
electromagnetism, whereas it is not altogether impossible that general relativity may not be 
a field theory
at all, but it may rather be a theory which describes the {\it thermodynamics} of space, time
and matter fields. Instead of attempting to quantize general relativity, we should try to 
find the microscopic constituents of spacetime, and to postulate for those constituents 
certain properties which imply, in the thermodynamical limit and at macroscopic length 
scales, Einstein's field equation as a sort of thermodynamical equation of state. This paper
may be regarded as an inquiry of a possibility to realize this kind of an approach to 
quantum gravity. Historically, it is very strange that although similar ideas were presented 
already by Misner, Thorne and Wheeler in their book \cite{kuusi}, and the intimate relationship between
gravity and thermodynamics was made obvious in the classic works of Hawking and Bekenstein 
on black hole thermodynamics \cite{kolme, nelja}, physicists have never made a really serious attempt to follow
these ideas systematically from the beginning to the end.

   Although our model has certain advantages in the sense that, in addition of predicting
some new effects of gravity, it reproduces everything we know with an (almost) absolute 
certainty about the properties of gravitation, the model also has certain problems. 
Conceptually, the most serious of the problems concerns an unclear role of {\it time}. 
Indeed, we brought, in Section 5, the concept of time into our model from "outside" at 
macroscopic length scales without any reference whatsoever to the microscopic properties 
of spacetime. More precisely, we associated with macroscopic regions of spacetime 
four-dimensional geometrical simplices, and on each simplex we had a flat Minkowski metric.
Since four-dimensional simplices played in our model a role similar to that of tangent 
spaces in classical general relativity, the role played by time in our model is similar to
the role it plays in general relativity: In general relativity the concept of time is 
brought in, when every point of spacetime is associated with a tangent space equipped with a
flat Minkowski metric. In our model time is a purely macroscopic concept, without any 
connection to the microphysics of spacetime. Indeed, the main problem of our model is that leaves 
this connection entirely unspecified. However, the problems concerning the concept of 
time are extremely difficult, and it seems that at this stage of research the only way to 
make progress may be to disregard those problems.

   Another problem concerns the cosmological constant. Our model predicts that the 
cosmological constant should be exactly zero, whereas recent observations seem to imply
that although the cosmological constant is extremely small, it is not exactly zero 
\cite{kolmekymmentayksi}. It would
be interesting to investigate, whether our model could be modified, in a natural manner, to
predict a non-zero value for the cosmological constant. 

   The main lesson one may learn from our model of spacetime is that it is indeed possible 
to construct a simple and mathematically reasonably precise quantum mechanical model of 
spacetime, which reproduces the "hard facts" of gravitational physics. Most likely our model
is still far away from a decent proposal for a proper quantum theory of gravity, but 
nevertheless the message it conveys is clear: Quantization of gravity may well be more 
simple than it has been thought so far.   

\begin{acknowledgments}
I thank Jorma Louko for reading the manuscript and suggesting improvements.
\end{acknowledgments}


\begin{thebibliography} {35}
\bibitem{kuusitoista} This idea has been explained in more details in Ref.[7]. See also the
references there.
\end{thebibliography}
\end{document}